\documentclass[fleqn,usenatbib]{mnras}
\usepackage{newtxtext,newtxmath}
\usepackage[T1]{fontenc}
\DeclareRobustCommand{\VAN}[3]{#2}
\let\VANthebibliography\thebibliography
\def\thebibliography{\DeclareRobustCommand{\VAN}[3]{##3}\VANthebibliography}
\usepackage{graphicx}	% Including figure files
\usepackage{amsmath}	% Advanced maths commands
\graphicspath{{Figures/}{../Figures/}}
\usepackage{xcolor}

\usepackage[normalem]{ulem}

\title[Model-independent constraints on $\Omega_m$]{Model-independent constraints on $\Omega_m$ and $H(z)$ from the link between geometry and growth}

\author[Ruiz-Zapatero et al.]{
Jaime Ruiz-Zapatero,$^{1}$\thanks{E-mail: jaime.ruiz-zapatero@physics.ox.ac.uk}
Carlos Garc\'ia-Garc\'ia,$^{1}$
David Alonso,$^{1}$
Pedro G. Ferreira,$^{1}$ 
\newauthor
~and Richard D.P. Grumitt $^{2}$
\\
$^{1}$ Department of Physics, University of Oxford, Denys Wilkinson Building, Keble Road, Oxford OX1 3RH, United Kingdom\\
$^{2}$Department of Astronomy, Tsinghua University, Beijing 100084, China}

\date{Accepted XXX. Received YYY; in original form ZZZ}
\pubyear{2021}

\begin{document}
\label{firstpage}
\maketitle

\begin{abstract}
We constrain the expansion history of the Universe and the cosmological matter density fraction in a model-independent way by exclusively  making use of the relationship between background and perturbations under a minimal set of assumptions.  We do so by employing a Gaussian process to model the expansion history of the Universe from present time to the recombination era. The expansion history and the cosmological matter density are then constrained using recent measurements from cosmic chronometers, Type-Ia supernovae, baryon acoustic oscillations, and redshift-space distortion data. Our results show that the evolution in the reconstructed expansion history is compatible with the \textit{Planck} 2018 prediction at all redshifts. The current data considered in this study can constrain a Gaussian process on $H(z)$ to an average $9.4 \%$ precision across redshift. We find $\Omega_m = 0.224 \pm 0.066$,  lower but statistically compatible with the \textit{Planck} 2018 cosmology. Finally, the combination of future DESI measurements with the CMB measurement considered in this work holds the promise of $8 \%$ average constraints on a model-independent expansion history as well as a five-fold tighter $\Omega_m$ constraint using the methodology developed in this work. 
\end{abstract}


%%%%%%%%%%%%%%%%%%%%%%%%%%%%%%%%%%%%%%%%%%%%%%%%%%

%%%%%%%%%%%%%%%%% BODY OF PAPER %%%%%%%%%%%%%%%%%%

\section{Introduction}
When observing and characterising the Universe on large scales, there are two broadly different, yet intertwined, types of observations \citep{1980lssu.book.....P}. In the first type of observation, one endeavours to constrain the expansion rate of the Universe at different times. This can be done by measuring the expansion rate itself or through a variety of cosmological distance measures: angular diameter distances, luminosity distances, standard sirens, etc. In particular, if $a(t)$ is the scale factor of the Universe at cosmic time $t$, the expansion rate at that time can be defined as $H[a(t)]={\dot a}/{a}$ (where the overdot is derivative with regards to cosmic time). Then, one can derive an expression for the radial comoving distance, $D_M$, by following a radial null-geodesic.
\begin{equation} \label{eq:distance}
    D_M \equiv \chi = \int_t^{t_0}\frac{dt'}{a(t')}=\int_0^{z}\frac{dz'}{H(z')}
    \, ,
\end{equation}
where we use unit $c=1$ throughout, $t_0$ is cosmic time today and $z$ is the cosmological redshift, $1+z=1/a$. Moreover, assuming zero spatial curvature, it is possible to derive expressions for the luminosity ($D_L$) and angular diameter distances ($D_A$) by multiplying and dividing Eqn. \ref{eq:distance} by $(1+z)$, respectively:
\begin{equation} \label{eq:distances}
    D_L = (1+z) D_M \quad \text{and} \quad
    D_A = D_M/(1+z).
\end{equation}
Thus, measurements of $D_L$ and $D_A$ at different redshifts can be used to reconstruct $H(z)$ over time.

In the second type of observation, one focuses on the rate with which structures undergo gravitational collapse on the expanding background. Under the assumptions that the bulk of matter in the recent past can be described by a presureless fluid and that neutrino effects can be neglected, the density contrast, $\delta\equiv\rho/{\bar \rho}-1$ (where ${\bar \rho}$ is the mean matter density of the Universe), obeys an evolution equation of the form:
\begin{eqnarray}
f'+f^2+\left(1+\frac{{\rm d}\ln aH}{{\rm d}\ln a}\right)f=\frac{3}{2}\Omega_m(z) \, ,  \label{eq:growth}
\end{eqnarray}
where the prime denotes a derivative with regards to $\ln a$, the growth rate is defined as $f\equiv {\rm d}\ln \delta/{\rm d}\ln a$ and $\Omega_m(z)$ is the fractional energy density in matter as a function of redshift. The latter quantity depends, through the Einstein field equations, on $H \equiv H(z)$ so that
\begin{eqnarray} \label{eq:jeans}    
\Omega_m(z)=\frac {\Omega_m H^2_0}{a^3H^2},
\end{eqnarray}
with $\Omega_m \equiv \Omega_m(0)$, which we use for ease of notation.
Thus, a measurements of the growth rate of structure at different redshifts (or times) can also be used to reconstruct $H(z)$ over time as well as the fractional matter density today, $\Omega_m$.

The challenge of modern cosmology has been to use a wide range of different cosmological observables to constrain $H(z)$ as a function of time or redshift (and, of course, $\Omega_m$) and, crucially, to pin down the underlying cosmological model which describes $H(z)$ in terms of a greatly reduced set of cosmological parameters. The frontrunner is the $\Lambda$ Cold Dark Matter ($\Lambda$CDM) model, whose parameters are now constrained to an unprecedented precision \citep{Planck}. However, interestingly, inconsistencies, or "tensions", have begun to emerge. Different probes are leading to different constraints on, for example, the Hubble constant, $H_0$ \citep[e.g.][]{riess_42, TRGB_H0, Maser_H0} or the density-weighted amplitude of fluctuations, $S_8 \equiv \sigma_8 \left(\Omega_m/0.3\right)^{0.5}$ \citep[e.g.][]{2007.15633, 2105.13549, 1607.00008, 2106.01208, K1K, 2105.09545, Garcia-Garcia21, 2021arXiv211109898W}, where $\sigma_8$ is the variance of $\delta$, in spheres of radius $8\,h^{-1}$Mpc.

In this paper we ask if it is possible to obtain meaningful, or precise, constraints on cosmological parameters with minimal assumptions about the cosmological model. To be more specific, we step back and try to find model-independent constraints on $H(z)$ and $\Omega_m$  from measurements of the expansion history itself, cosmological distances and the growth rate. This allows us to contrast the constraining power of these two very different sets  of observables, to explore how combining them improves constraints and, most importantly, how much constraints improve once one assumes a cosmological model.
The hope is that understanding this process will shed light on the analysis of theories that go beyond  $\Lambda$CDM, but also may have bearing on the current inconsistencies in parameter constraints. We will implement our model-independent approach using Gaussian processes \citep{gp_book, Seikel_12}. Gaussian processes allow us to  reconstruct a well-defined distribution of histories of $H(z)$. The data then allows us to constrain the parameters of this distribution and, in doing so, tells us at what redshifts $H(z)$ is well determined and at what redshifts it is determined poorly.

The literature already hosts a number of examples of the possible uses of Gaussian processes to test certain aspects of the present cosmological paradigm in model-agnostic ways. For example,  one can find tests for the model dependence of the $H_0$ \citep{1802.01505, 1907.10813, Liao_H_gp, Bonilla_H0_gp} and $S_8$ \citep{S8_gp, Huillier18} tensions, for the non-zero curvature of space-time \citep{2011.11645, 2007.05714, Shafieloo18}, as well as tests for the density and the equation of state of dark energy \citep{1902.0942, 1806.02981}.  More closely related to the topic of this work, in \citet{2105.01613} Gaussian processes were used to study the statistical correlations between the expansion history, cosmological distances, and the linear growth rate without appealing to the physical relationships between the three functions. However, the clearest precedent of the methodology used here is \citet{1911.12076} who already employ Gaussian processes to obtain model-independent constraints for $\Omega_m$ and $\sigma_{8}$ based on the relationship between the expansion history and the linear growth rate.  The methodology developed in the present paper expands and improves many aspects of their analysis. First, our Gaussian process extends all the way to recombination. This allows us to solve the Jeans equation (Eqn. \ref{eq:growth}) without using fitting formulas. Moreover, we also employ cosmological distances to constrain the evolution of the expansion history. This, in combination with the extended range of the Gaussian process, allows us to use the position of the first acoustic peak of the CMB temperature power spectrum to constrain expansion history far into the past. Finally, and most importantly, we sample our Gaussian process simultaneously with the cosmological parameters, allowing us to observe potential correlations between the two.

The structure of this paper is as follows: in Section~\ref{Sect: Methodology} we present the methodology of this work regarding the use of Gaussian Process to compute predictions for cosmological observables. In Section~\ref{Sect: Observables and data sets} we describe the cosmological observables from which we employ data and motivate their use in the context of this work. We present our results in Section~\ref{Sect: Results} and discuss the implications of our work in Section ~\ref{Sect: Conclusions}.

\section{Methodology} \label{Sect: Methodology}

Throughout this paper we will be working with the Eqns. \ref{eq:distance} and \ref{eq:growth} to connect the time evolution of $H(z)$ with current and future observations. In the case of direct measurements of $H(z)$, the expansion history can be trivially matched to observations without the need to apply any transformation.  In the case of distances, we can see from Eqn. \ref{eq:distance} that it is possible to generate predictions for the observables from any choice of $H(z)$ by performing the relevant integral. In the case of growth (Eqn. \ref{eq:growth}), we are faced with the problem that most observations do not report $f$ but the combined quantity $ f\!\sigma_{8}$. Thus, it is convenient to rewrite Eqn. \ref{eq:growth} in terms of the latter. This can be done as follows: assuming that perturbations grow in a self-similar manner, we can  define $\delta(t,{\bf x})\equiv D(t)\delta_0({\bf x})$, where $\delta_0$ is the density contrast today. Then, we can rewrite Eqn. \ref{eq:growth} as
\begin{equation} \label{eq:jeans_D}
    \frac{d}{da}\left(a^2H\frac{dD}{d\log a}\right)=\frac{3}{2}\Omega_m(a)\,a\,H\,D \, .
\end{equation}
Defining $y\equiv a^2E\frac{dD}{d\log a}$, with $E\equiv H/H_0$, and switching to the integration variable $s\equiv\log(1+z)$, this equation can be written in terms of a system of coupled  first-order equations:

\begin{equation} \label{eq:Jeans_sys_D}
    \frac{dy}{ds}=-\frac{3}{2}\frac{\Omega_m}{a E}D \, ,\hspace{12pt} 
    \frac{dD}{ds}=-\frac{y}{a^2E}.
\end{equation}

Then, it is possible to transform the quantities $y(s)$ and $D(s)$ into $f\!\sigma_{8}(s)$ and $\sigma_{8}(s)$ by applying the following transformation
\begin{equation} \label{eq:D_to_s8}
    \sigma_{8}(s) = \sigma_{8}(0) \frac{D(s)}{D(0)}\,,\hspace{12pt}
    f\!\sigma_{8}(s) = \frac{y(s) \, \sigma_{8}(0)}{E(s)D(0)}e^{2s}\, .
\end{equation}
The thrust of this paper is to keep the analysis as model-independent as possible. Yet, as we can see, it is still possible to extract information about some of the cosmological parameters. For a start, combining the information from distances and growth allows us to constrain $\Omega_m$. But we also have, automatically, $H_0\equiv H(z=0)$ and, as we just saw, we can calculate $\sigma_8\equiv\sigma_8(z=0)$ (or $S_8$ as a derived parameter).

In this paper we will model the time (or redshift) dependent Hubble rate, $H(z)$, as a Gaussian Process spanning over the redshift range $0<z<1100$. A Gaussian process (GP) is a collection of random variables, any finite number of which have a joint Gaussian distribution \citep{gp_book}. A GP $g(\textbf{x})$, where $\boldsymbol{x}$ is a collection of random variables, is fully specified by a mean function $m(\textbf{x}) \equiv \mathcal{E}[g(\textbf{x})]$, where $\mathcal{E}[\cdots]$ is the expectation value over the ensemble,  and a covariance function $k(\textbf{x}, \textbf{x'}) \equiv \mathcal{E}[(g(\textbf{x})-m(\textbf{x}))(g(\textbf{x'})-m(\textbf{x'}))^T]$. In combination, the mean and covariance functions determine the statistical properties of the ensemble of random variables which defines the family of shapes that the GP can take. 

Thus, GP's can be used as agnostic function-space priors which, in combination with a likelihood for the observed data $\mathcal{L}(\textbf{y}| g(\textbf{x}), \boldsymbol{\sigma})$, where $\textbf{y}$ is a set of data points with a set of errors $\boldsymbol{\sigma}$,  define a regression model. Then, observations can be used to inform the GP posterior (i.e. the statistical properties of the set of random variables) $\mathcal{P}(g(\textbf{x})| \textbf{y},\boldsymbol{\sigma})$, which determines the subset of functions most consistent with the data.

 Since the Hubble rate is generally regarded as a monotonically increasing function, it is important to define a non-zero mean for the GP. This prevents the GP from simply fitting the long range upwards trend of $H(z)$ while washing out interesting local features \citep{Shafieloo_gp}. In this paper, we will define the mean of the GP in terms of a fiducial $\Lambda$CDM prediction:
\begin{equation} \label{H_bar}
    \overline{H}(z) \equiv  \overline{H_0} \sqrt{\overline{\Omega}_m(1+z)^3+\overline{\Omega}_r(1+z)^4+ \overline{\Omega}_{\Lambda}} \, ,
\end{equation}
where $\overline{\Omega}_m$,  $\overline{\Omega}_r$ and $\overline{ \Omega}_{\Lambda}$ stand for the present values of the cosmological densities of matter, radiation and dark energy , respectively,
as given by the \textit{Planck} 2018 TTTEEE+lowE cosmology (see Tab. \ref{tab:fiducial}). In other words, they are completely fixed and are not variables.

In order to prevent this (ultimately arbitrary) choice of mean from biasing our constraints on the cosmological parameters, we define a free amplitude parameter $A_0$ that multiplies the mean of the GP such that
\begin{equation} \label{H_mean}
    H_m(z) \equiv A_0 \, \overline{H}(z) \, .
\end{equation}
Simply put, if the data is somewhat away from what one might expect from the fiducial $\Lambda$CDM background, the Gaussian process will have to soak up the large scale differences, or trends, in detriment to the small scale, local features. Moreover, in those regions where data is sparse, the GPs constraints are highly dominated by their prior; i.e. the chosen mean. If such mean systematically falls beneath or above the data, it can lead to spurious trends in the final results. In App. \ref{app:validation}, we show these effects can substantially bias the constraints we obtain in the absence of $A_0$. For these reasons, we advise against common practices such as employing a constant as a mean for a GP modelling the recent expansion history or extrapolating GP results to regions where they become dominated by the choice of mean and covariance function. 

\begin{table} 
    \caption{ Values of the cosmological parameters used to define the fiducial cosmology. Values based of \textit{Planck} 2018 TTTEEE+lowE cosmology \citep{Planck}}
    \centering
    \begin{tabular}{|p{1.5 cm}|p{1.5cm}}
    \hline
    Parameter & value   \\
    \hline
    $\overline{\Omega}_{\rm{m}}$   &  0.316  \\
    $\overline{\Omega}_{\Lambda}$ & 0.683 \\
    $\overline{\Omega}_r$ &  $9.245 10^{-5}$ \\
    $\overline{H}_0$ & 67.27 \\
\end{tabular}
\centering
\label{tab:fiducial}
\end{table}

The Hubble rate must also be a continuous and smooth function. Therefore it is important that the covariance function of the GP reflect such properties \citep{Dialektopoulos21}. Ultimately, we chose a a square exponential covariance function to model the correlations between the different nodes of the GP. This decision was made based on the fact that the square exponential is computationally inexpensive and infinitely differentiable kernel appropriate for modelling smooth fluctuations around the mean of the GP. Mathematically, the square exponential covariance function is defined as
\begin{eqnarray} 
k\left[g(\boldsymbol{x}),g(\boldsymbol{x}')\right]=\eta^2\exp\left[\frac{-|g(\boldsymbol{x})-g(\boldsymbol{x}')|^2}{2l^2} \right] \, ,  \label{eq:QuadExp}
\end{eqnarray}
where $\eta$ is the amplitude of the oscillations around the mean and $l$ is the correlation length between the GP nodes. The parameters $\eta$ and $l$ are known as hyperparameters since they constrain the possible values that the nodes; i.e. the parameters the GP, can take. In other words, the hyperparameters dictate the family of functions available to the GP.  In addition to this, a white noise term with amplitude $10^{-3}$ was added to the covariance function to ensure numerical stability.   

Sampling over GPs is extremely efficient when the model for the observed data is linearly related to the GP and the likelihood is Gaussian. In this case, it can be shown that the posterior distribution also takes the form of a GP for which the mean and covariance functions can be calculated analytically by marginalizing over the nodes of the original GP \citep{gp_book}. However, when the observed data is not linearly related to the GP, the posterior distribution becomes non-Gaussian. Thus, each node of the GP; i.e. each dimension of the multivariate Gaussian distribution, must be treated as a newly added degree of freedom in the sampling process (i.e. a new parameter). 
 
\begin{figure*} 
\centering
    \includegraphics[width=.65\linewidth]{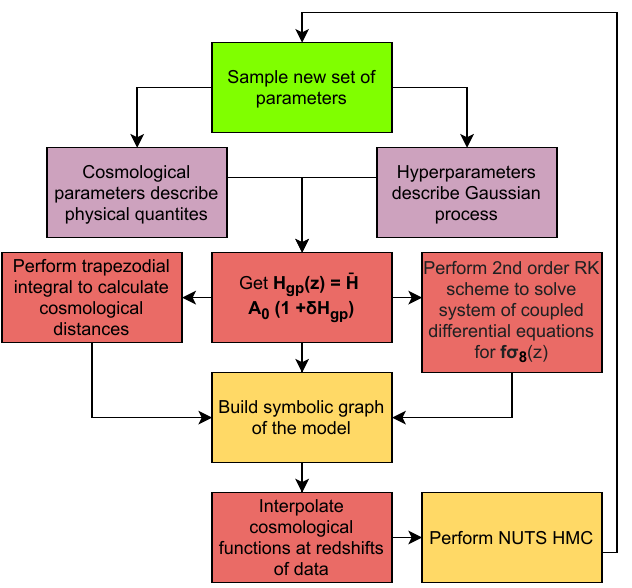} 
    \caption{Shows a schematic representation of the data analysis pipeline.}
    \label{fig:code}
\end{figure*} 

Exploring such a vast parameter space is effectively unfeasible with traditional non-gradient-based sampling algorithms and requires more sophisticated samplers. In this work we employ the No U-turn Sampler (NUTS) \citep{NUTS}, a self-tuning variant of Hamiltonian Monte Carlo (HMC) \citep{HMC, Betancourt17} which simulates Hamiltonian trajectories through the parameter space to generate efficient proposal steps. This allows us to handle hundreds or thousands of parameters during the inference process efficiently. 
We employ the NUTS implementation of the python package \texttt{Pymc3} \citep{Pymc3} which makes use of the tensor package \texttt{Theano} \citep{Theano} as a back-end to build a symbolic graph of the model used to fit the data. This graph is then used to perform automatic-differentiation \citep{Autodiff18} to obtain the gradient of the regression model needed for HMC. Through out the likelihood we make use of simple integration schemes; e.g. Runge-Kutta, trapezoidal rule, on which \textsc{Theano} can apply automatic-differentiation to make the likelihood differentiable in all parameters.

In the case of distances (see Eqn. \ref{eq:distance}), the integral over $1/H_{\mathcal{GP}}$ was performed using the trapezoidal rule with $s\equiv \log(1+z)$ as the integration variable. In the case of the linear growth rate, the system of differential equations shown in Eqns. \ref{eq:Jeans_sys_D} was solved for $y(s)$ and $D(s)$ employing a second order Runge-Kutta scheme, integrating over the redshift range $0 < z < 1100$ from the past into the present with initial conditions $D(z=1100)=a(z=1100)$ and $y(z=1100)=a(z=1100)^3E(z=1100)$. Note that these initial conditions are only strictly correct for a purely matter-dominated universe, which is not accurate at $z\sim1000$. However, the impact of this assumption is negligible by comparison to current uncertainties on growth measurements. In combination, these methods allow us to obtain sub-percentage accuracy through out the entire redshift range of the GP with respect the output of cosmological CLASS \citep{Class}, which performs the full numerical calculation.

In order to cover the redshift range $0 < z < 1100$ while keeping the numerical errors under control, we sampled the GP evenly in $s$ as opposed to redshift itself. This variable concentrates most of the GP samples at low redshift where most of our data lays while offering great numerical accuracy \citep{1710.0423}. This allowed us to cover our desired redshift range with only $200$ nodes; i.e. with a 200-dimensional GP. 

In addition to the $200$ parameters associated with the GP, we also sample over the cosmological parameters present in Eqn. \ref{eq:growth}; i.e. $\Omega_m$ and $\sigma_{8}$, as well as the GP hyperparameters $\eta$ and $l$ described in Eqn. \ref{eq:QuadExp}. We define the GP not over the Hubble rate itself but over as relative deviations from the $\Lambda$CDM background 
\begin{eqnarray}
  H_{\mathcal{GP}} \equiv H_m(s) [1 + \delta H_{\mathcal{GP}}(s)], \nonumber
\end{eqnarray}
where $\delta H_{\mathcal{GP}}$ is a GP centred at zero. This approach has the added benefit of normalizing the amplitude of the oscillations of the GP with respect the mean making the sampling of $\eta$ far more efficient by virtue of reducing its potential range of values. Note that the correlation length scale $l$ is defined in units of $s$ as opposed to $z$. We also marginalize over the absolute magnitude of the supernovae $M$ (See Sect. \ref{subsec: Supernovae} for details), the scale of the sound horizon $r_s$, and the amplitude of $H_m$, $A_0$.

It is important to emphasise that we perform a fully Bayesian inference over the GP, as opposed to what it is known as an empirical Bayesian analysis in which first, the marginal likelihood of the GP hyperparameters is maximized; and then, keeping their values fixed, the conditional posterior over the GP is inferred. While for large sets of data the output of the two approaches converges to the same results, when only sparse data is available the fact that the empirical Bayesian analysis does not account for the full posterior volume of the hyperparameters  can lead to an under-accounting of uncertainties.

In order to establish the reliability of the methodology developed in this work for cosmological inference, and the robustness of the fiducial results, we also considered a series of alternative analyses (see \citet{Colgain21} for a discussion of different GP implementations). In the first test, the hyperparameters of the GP ($\eta$, $l$) were kept fixed. We will label results from this analysis as "Fixed HP". In the second test, instead of sampling over the $A_0$ parameter, we performed a two-step analysis. In the first step, a standard $\Lambda$CDM model was fit to the data. Based on this best-fit $\Lambda$CDM model, a expansion history was derived. Then, in the second step, the best-fit expansion history was used as the mean of the GP during the analysis while fixing $A_0 = 1$. We will label results from this analysis as "Two-Steps". 

\begin{table} 
\caption{Sampled parameters and their priors in the standard analysis (first column), the fixed hyperparameters analysis (second column) and the Two-Steps analysis (third column). }
\begin{tabular}{ |p{1cm}|p{1.75cm}|p{1.75cm}|p{1.75cm}|p{1.75cm}|p{1.75cm}}
 \hline
  Parameter & Fiducial & Fixed HP & Two-Steps  \\
    \hline
    $\eta$ & $\mathcal{N}_{1/2}(0,0.2)$  & 0.2 & $\mathcal{N}_{1/2}(0,0.2)$  \\
    $l$ & $U(0.01, 6)$  & 1.0 & $U(0.01, 6)$   \\
    \hline
    $\Omega_m$ & $U(0,1)$ & $U(0,1)$ & $U(0,1)$ \\
    $\sigma_{8}$ & $\mathcal{N}(0.8, 0.5)$ & $\mathcal{N}(0.8, 0.5)$ & $\mathcal{N}(0.8, 0.5)$   \\
    \hline
    $A_0$ & $U(0.8, 1.2)$ & $U(0.8, 1.2)$ & 1 \\
    $M$ & $\mathcal{N}(-19.2, 1)$ & $\mathcal{N}(-19.2, 1)$ & $\mathcal{N}(-19.2, 1)$   \\
    $r_s$ & $\mathcal{N}(150, 5)$ & $\mathcal{N}(150, 5)$ & $\mathcal{N}(150, 5)$ \\
\end{tabular}
\label{tab:prior}
\end{table}

A summary of the priors used in the fiducial and alternative analyses can be found in Tab. \ref{tab:prior}. As a general rule, when performing Bayesian inference one should avoid broad, uniform priors \citep{Gelman17}. This is mainly due to the fact that they do not accurately represent the prior knowledge, put a lot of posterior mass in unlikely values and introduce hard boundaries which can be difficult to motivate. This last consideration is particularly important when using \textsc{HMC} since sharp edges in parameter space can lead to the sampling process becoming inefficient. Following this principle, we employed a half-Gaussian distribution with zero mean and standard deviation $0.2$ as prior distribution for the hyperparameter $\eta$ to down-weight extreme deviations (i.e. $20\%$ and above, well within the observational errors on $H(z)$) from the chosen GP mean without introducing any unnecessary hard boundaries. In the case of $l$ we used a uniform distribution between $0.01 < l < 6$ for the following reasons. First, there is no reason to down-weight long correlation modes against short ones or vice versa. Second, looking at Eqn. \ref{eq:QuadExp}, it is possible to see that as $l \rightarrow 0$ the value of the covariance function approaches zero regardless of the value of $\eta$. This opens a vast a volume in the parameter space which can lead to an inefficient sampling. Third, the likelihood function becomes flat at such small scales since there is no information in the data to constrain those small scale modes. Forth, the expansion rate is expected to have some degree of temporal correlations, and therefore the limit $l=0$ must be excluded. The parameter $\Omega_m$ was sampled from a uniform distribution between $0$ and $1$ to enforce the physical boundaries on the allowed values for the cosmological matter fraction. In the case of $\sigma_8$ for which there is no physically-motivated upper limit, we employ a better behaved Gaussian prior centered at $0.8$ and with $0.5$ standard deviation. Given the degeneracy of $A_0$ with the GP parameters, the parameter was sampled from a uniform distribution between $0.8$ and $1.2$, which amply encompasses the current discrepancy in $H_0$ between CMB data and local measurements. The supernova absolute magnitude parameter $M$ was sampled from a Gaussian distribution centered at value found by the SH0ES collaboration for the parameter $M = -19.2$, with a standard deviation of $1$. While the mean value corresponds to a local expansion rate of $H_0 \simeq 74.0$ km/s/Mpc, the standard deviation ensures that all values in the range $50 < H_0 < 100$ km/s/Mpc fall within the $1\sigma$ region of the prior distribution. Finally, the sound horizon scale, $r_s$,  was sampled from a Gaussian distribution centered at $150$ Mpc with a standard deviation of $5$ Mpc. We do this instead of computing $r_s$ from the expansion history in combination with the BBN prior to allow for largest deviations from the fiducial cosmology. This was needed to recover the test cosmologies of App. \ref{app:validation}.

%%%% Observables and data
\section{Observables and Data sets} \label{Sect: Observables and data sets}
In order to make the most of the modelling freedom offered by GPs we consider as much data as possible. In this work we use a combination of several different probes that together account for 91 data points for a variety of cosmological observables. A summary of the data used in this work can be found in Tab. \ref{tab:data}.

\begin{figure*} 
\centering
    \includegraphics[width=\linewidth]{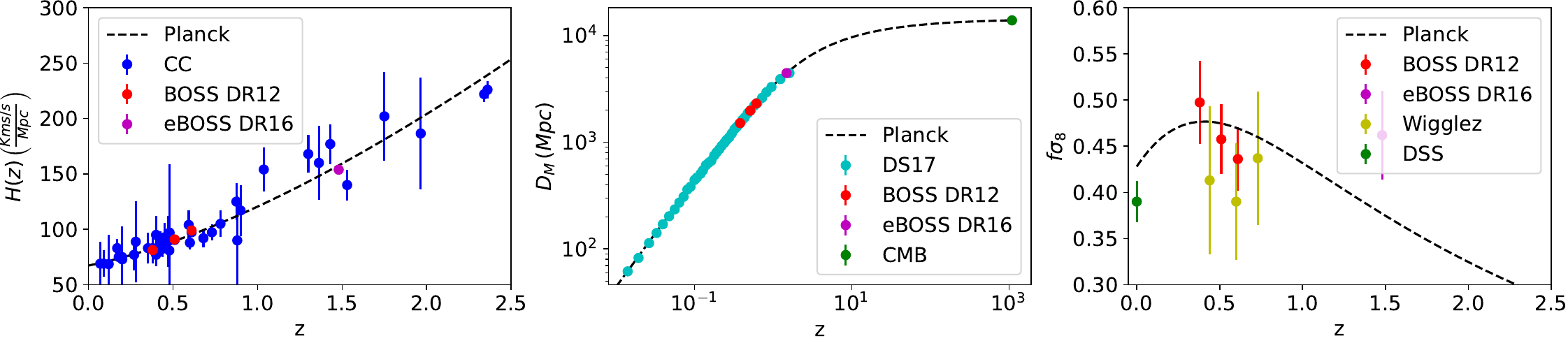} 
    \caption{Data points from the different surveys used in this work across redshift for the three cosmological functions of interest $H(z)$, $D_M$ and $f \! \sigma_8$.}
    \label{fig:data}
\end{figure*} 

\begin{table*} 
\centering
\caption{Data sets used in our analysis, listing the probe, the redshift range of the probe, the choice of observable and the size of the data vector. See Fig. \ref{fig:data} for a pictorial representation. }
\begin{tabular}{|p{5.5cm}|p{2cm}|p{1.5cm}|p{0.7cm}|p{.7cm}|p{0.7cm}|p{1cm} |  }
 \hline
Data set &  Probe & Redshifts &\multicolumn{3}{|c|}{Observable} & Data Points\\
&  & & $H(z)$ & $D_M(z)$ & $f \! \sigma_8 $ & \\
 \hline
 CCs \citep{2011.11645} &  Cosmic Chronometers & 0.07 - 2.36 & $\checkmark$ & $\times$ & $\times$ & 35\\
 
Pantheon DS17\citep{1710.00845} & SNe Ia & 0.38 - 0.61 & $\times$ & $\checkmark$ & $\times$ & 40\\
 
 BOSS DR12 \citep{1607.03155} & BAO+RSD & 0.38 - 0.61 & $\checkmark$ & $\checkmark$ & $\checkmark$ & $3\times3$\\

 eBOSS DR16 \citep{2007.08991} & BAO+RSD & 1.48 & $\checkmark$ & $\times$ & $\times$  & $1\times3$ \\ 
 
 Wigglez \citep{1204.3674} & RSD & 0.44 - 0.73 & $\times$ & $\times$ & $\checkmark$  & $1\times3$ \\
 
 DSS \citep{2105.05185} & RSD & 0 & $\times$ & $\times$ & $\checkmark$  & 1 \\
 
 \textit{Planck} 2018  \citep{Planck} & CMB & 1090.30 & $\times$ & $\checkmark$ & $\times$  & 1 \\
 \hline
 DESI \citep{Font-Ribera14} & BAO+RSD & 0.15 - 1.85 & $\checkmark$ & $\checkmark$ & $\checkmark$ & $3\times18$\\

\end{tabular}
\label{tab:data}
\end{table*}

\subsection{Cosmic Chronometers}

Cosmic Chronometers (CCs) are tracers of the evolution of the differential age of the Universe as a function of redshift. Since $H(z) \equiv \Dot{a}/a = -(dt/dz)/(1+z)$ a measurement of dt/dz directly yields the expansion rate \citep{0106145}.  By measuring the age difference between two ensembles of passively evolving galaxies at different redshifts, one can determine the derivative of redshift with respect to cosmic time, $dz/dt$. Massive, early, passively-evolving galaxies have been found to be very good tracers in this sense \citep{Cimatti06, Thomas11, Moresco15, Moresco18, Moresco20} and have been used extensively over the past two decades to measure $H(z)$ up to $z \approx 2$. In this work we make use the $H(z)$ measurements from CCs summarized in Table 1 of \citet{2011.11645}.

\subsection{Supernovae}
\label{subsec: Supernovae}
Type Ia supernovae (SNe Ia) — titanic explosions of white dwarfs in multi-star systems  \citep{Hoyle60, Colgate69}, are highly prized observations in cosmology due to their capacity to act as standard candles \citep{9907052, 0012376}.

However, SNe Ia by themselves can only inform their relative distance to one another and need to be calibrated with nearby SNe Ia of known redshift and luminosity distance $D_L$ to obtain absolute distances. Thus, SNe Ia inform the relationship
\begin{equation} \label{eq: light_curve}
    \mu(z) = 5 \log_{10} D_L(z) + 25 + M \,, 
\end{equation}
known as the luminosity distance modulus where $M$ is the calibrator known as the absolute magnitude of the SNe Ia. Therefore, once $M$ has been determined SNe Ia luminosity distance measurements can be used to inform the Hubble rate.

In this work, we fit the compressed data vector of the Pantheon sample, known as "DS17", composed of 40 measurements of the distance modulus (See Eqn. \ref{eq: light_curve}) in the range $0.15\leq z\leq1.615$ \citep{1710.00845}. The original Pantheon sample is composed of the optical light curves and redshifts for 365 spectroscopically confirmed Type Ia supernovae (SNe Ia) discovered by the Pan-STARRS1 (PS1) Medium Deep Survey combined with the subset of 279 PS1 SN Ia ($0.03 < z < 0.68$) with useful distance estimates of SN Ia from SDSS, SNLS, various low-z and HST samples to form the largest combined sample of SN Ia consisting of a total of 1048 SN Ia ranging from $0.01 < z < 2.3$ \citep{1710.00845}. 

In the light of recent works in the literature questioning the accuracy of the absolute calibration of important sectors of Pantheon sample \citep{H0_ofer}, we marginalize over the absolute magnitude of the supernovae as opposed to fixing its value. Due to the degeneracy between $M$ and $H_0$, this is equivalent to fitting the expansion rate, $E(z)= H(z)/H_0$, as opposed to the Hubble rate, $H(z)$.

\subsection{Baryon Acoustic Oscillations}
Baryon acoustic oscillations (BAOs) enhance matter overdensities  at a characteristic physical separation scale which corresponds to the size of the sound horizon at the end of the drag epoch, $r_{\rm{s}}(z_{\rm d})$ \citep{Peebles70, 0110414}. The sound horizon is defined as the distance a pressure wave can travel from its time of emission in the very early Universe up to a given redshift. This can be expressed as 
\begin{equation} \label{eq:rsound}
    r_{\rm{s}}(z)= \int^{\infty}_z \frac{c_{\rm{s}}  \, {\rm d}z'}{H(z')} \;,
\end{equation}
where $c_{\rm s}$ denotes the speed of sound, and where $H(z)$ is the expansion rate at redshift $z$. The end of the drag epoch is defined as the time when photon pressure can no longer prevent gravitational instability in baryons around $z \sim 1020$ \citep{0803.0547}. 

The BAO feature can be measured in the directions parallel and perpendicular to the line of sight. Perpendicular to the line of sight, the BAO feature informs the trigonometric relationship
\begin{equation}\label{eq:acoustic}
    \theta = \frac{r_{\rm{s}}(z_{\rm d})}{D_M(z)} \;,
\end{equation}
where $\theta$ is the angle under which the scale of the sound horizon is observed.  Parallel to the line of sight, the BAO feature informs the relationship $ \Delta z = H(z) r_{\rm{s}}(z_{\rm{d}})$ which can be used to constrain the expansion history of the Universe directly.

In this work, we make use of the twelfth data release of the galaxy clustering data set of the Baryon Oscillation Spectroscopic Survey (BOSS DR12) which forms part of the Sloan Digital Sky Survey (SDSS) III. BOSS DR12 comprises 1.2 million galaxies over an area of 9329 deg$^2$ and volume of 18.7 Gpc$^3$, divided into three partially overlapping redshift slices centred at effective redshifts 0.38, 0.51, and 0.61.

We fit the Alcock-Paczynski (AP) parameters  $\alpha_\parallel$ and $\alpha_\perp$ as reported by the BOSS DR12 data products
 \begin{equation} \label{eq:alphas_B}
    \frac{[H(z)]_{\rm{fid}}}{\alpha_\parallel} = \frac{H(z) [r_{\rm{s}}(z_{\rm{d}})]_{\rm{fid}} }{ r_{\rm{s}}(z_{\rm{d}})}, \quad
    \alpha_\perp [\chi(z)]_{\rm{fid}} = \frac{\chi / r_{\rm{s}}(z_{\rm{d}})}{[r_{\rm{s}}(z_{\rm{d}})]_{\rm{fid}}} ,
\end{equation}
from the reconstruction of the BAO feature at the three different redshift bins where $[r_{\rm{s}}(z_{\rm{d}})]_{\rm{fid}} = 147.78$ Mpc is the scale of the sound horizon at drag epoch as given by the fiducial cosmology used for the reconstruction. $[H(z)]_{\rm{fid}}$ and $[\chi(z)]_{\rm{fid}}$ are the corresponding Hubble parameter and comoving radial distance for the fiducial cosmology, respectively. 

In addition to this, we employ the  anisotropic clustering of quasars in the sixteenth data release of the extended Baryon Oscillation Spectroscopic Survey (eBOSS DR16 \citealp{1703.00052}), which forms part of the Sloan Digital Sky Survey (SDSS) IV  \citep{1508.04473}. The eBOSS DR16 catalog contains 343,708 quasars  between $0.8 < z < 2.2$, from which BAO and RSD meausurements are obtained at an effective redshift of $z_{\rm eff} = 1.48$ \citep{2007.08998}.  We use the results from the configuration space analysis performed by measuring the two-point correlation function and decomposing it using the Legendre polynomials.  Similarly to BOSS DR12, the BAO signal is measured both parallel and perpendicular to the line of sight. This allows for the measurement of the geometrical relationships $D_H(z_{\rm eff})/[r_{\rm d}]_{\rm fid}$ and $D_M(z_{\rm eff})/[r_{\rm d}]_{\rm fid}$ respectively, where $D_H(z) \equiv c/H(z)$ and $[r_{\rm d}]_{\rm fid} = 147.3$.

Finally, we make use of the \textit{Planck} 2018 measurement of the BAO angular scale $\theta_* = \frac{D_{\rm{A}}(z_*)}{r_{\rm{s}}(z_*)}$, where $z_* \sim 1100$ is the redshift of the last scattering surface. We use the \textit{Planck} measurement from the temperature and polarization maps denoted as TTTEEE + lowE.

\subsection{Redshift Space Distortions}
Redshift space distortions (RSDs) are modifications to the observed redshift of a given object caused by its radial peculiar velocity \citep{kaiser87}. RSDs are caused by deviations
from the Hubble flow that are gravitationally induced by inhomogeneities in the gravitational potential of the surrounding matter distribution. On large, linear scales RSDs are dominated by the infall towards overdense structures, known as the Kaiser effect \citep{Hamilton98}. Clustering two-point statistics as a function of transverse and line-of-sight separation are sensitive to the quantity $f \! \sigma_8$ via RSDs \citep{1607.03150}.
Alternatively, SNe Ia themselves can be used as a probe of the local velocity field \citep{2105.05185, 9908237, 1605.01765}, which can also be used to measure this parameter.

We use the three measurements of  $f\sigma_{\rm{8}}(z)$ from RSD from the BOSS DR12 data, obtained using the anisotropic clustering of the pre-reconstruction density field \citep{1607.03155}. We also include the value of $f \! \sigma_8 (z_{\rm eff})$ measured from the BOSS DR16 quasar sample. The BAO and RSD measurements of both datasets are extracted from the same set of observations. As such, they are not statistically independent, and we account for their full covariance matrix in our analysis \citep{1607.03155,2007.08998}. In addition to these, we use the $f \! \sigma_8$ measurements reported by the WiggleZ Dark Energy Survey at the Australian Astronomical Observatory \citep{Drinkwater10} at redshifts 0.44, 0.60 and 0.73.

Finally, we also use the value of $f \! \sigma_8 (z=0)$ derived from the measured peculiar velocities of the Democratic Samples of Supernovae \citep[DSS]{DSS}, the largest  catalog used to study bulk flows in the nearby Universe, compiled of 775 low-redshift Type Ia and II supernovae (SNe Ia $\&$ II).

\subsection{Synthetic Stage-IV Data}
\label{sec: forecasts}
In addition to the previously discussed observables and data sets, we also produce forecasts for future experiments, focusing on the Dark Energy Spectroscopic Instrument (DESI). DESI is a galaxy and quasar redshift survey currently taking data from the Mayall 4 meter telescope at Kitt Peak National Observatory. The baseline area is 14000 sq. deg. We use the forecast errors for the measurements of the Hubble rate, $H(z)$; the angular diameter distance, $D_A(z)$; and the growth rate measurement $f \! \sigma_8$ reported by \citet{Font-Ribera14} to create a set synthetic measurements for $H(z)$, $D_A(z)$, and $f \! \sigma_8$ at an array of 18 redshifts from 0.15 to 1.85. The synthetic measurements were generated using the best-fit \textit{Planck} 2018 cosmology, including measurement noise following the statistical uncertainties reported by \citet{Font-Ribera14}.

%%% RESULTS
\section{Results} \label{Sect: Results}

\begin{table} 
\caption{One dimensional constraints on the hyperparameter $\eta$ (first column) and the mean variance of $\delta H(z)$ (second column) for different analyses.}
\begin{tabular}{|p{3.2cm}|p{2cm}|p{2cm}|}
 \hline
  Analysis & $\eta$ & $\overline{\sigma}(\delta H(s))$ \\
    \hline
    All data ($\Lambda$CDM) & $0 \pm 0$ & $0.011 \pm 0.003$ \\
    DESI+CMB ($\Lambda$CDM) & $0 \pm 0$ & $0.002 \pm 0.001$ \\
    \hline
    All data & $0.113 \pm 0.075$ & $0.094 \pm 0.004$ \\
    All data (Fixed HP) & $0.2$ & $0.2$ \\
    All data (Two-Steps) & $0.044 \pm 0.039$ & $0.025 \pm 0.01$ \\
    No CMB & $0.142 \pm 0.096$ & $0.125 \pm 0.016$ \\
    No DSS & $0.111 \pm 0.074$ & $0.091 \pm 0.005$ \\
    Growth data & $0.147 \pm 0.114$ & $0.183 \pm 0.003$ \\
    Geometry data & $0.122 \pm 0.077$ & $0.098 \pm 0.005$ \\
    \hline
    DESI $+$ CMB & $0.085 \pm 0.074$ & $0.08 \pm 0.004$ \\
\end{tabular}
\label{tab:variance}
\end{table}

In this section we present the results of our analysis, including both the constraints on expansion history and cosmological parameters from current data, and forecasts for ongoing and future surveys. 

\subsection{Validity of the methodology}

Before discussing the results of our analysis, we must first verify its reliability. Our methodology produces two types of results: constraints on the GP used to model the expansion history, and on the cosmological parameters (mainly $\Omega_M$) involved in solving the Jeans equation (Eqn. \ref{eq:growth}). We thus report results on both aspects. In App. \ref{app: Alternative analyses} we present the results of the alternative analyses described at the end of Sect. \ref{Sect: Methodology} and show that, despite the different treatments of the GP, they produce statistically compatible constraints for the cosmological functions as well as for the cosmological parameters, and similar uncertainties on both.  In App. \ref{app:validation}, we considered five different cosmologies, each progressively more discrepant with our fiducial cosmology (\textit{Planck} 2018), and generated mock data based on each them. We then showed that our methodology recovers the input cosmologies within statistical uncertainty. In addition to this, in App. \ref{app: Using HMC}, we discuss the specific \textsc{NUTS} set up used to produce our results. These settings allow us to achieve a Gelman-Rubin convergence test~\citep{GelmanRubin} value of $R -1 < 0.01$ for all the combinations of data considered.

%%%% COSMOLOGICAL FUNCTIONS
\subsection{Current constraints}
\subsubsection{Cosmological functions}

\begin{figure} 
\centering
    \includegraphics[width=\linewidth]{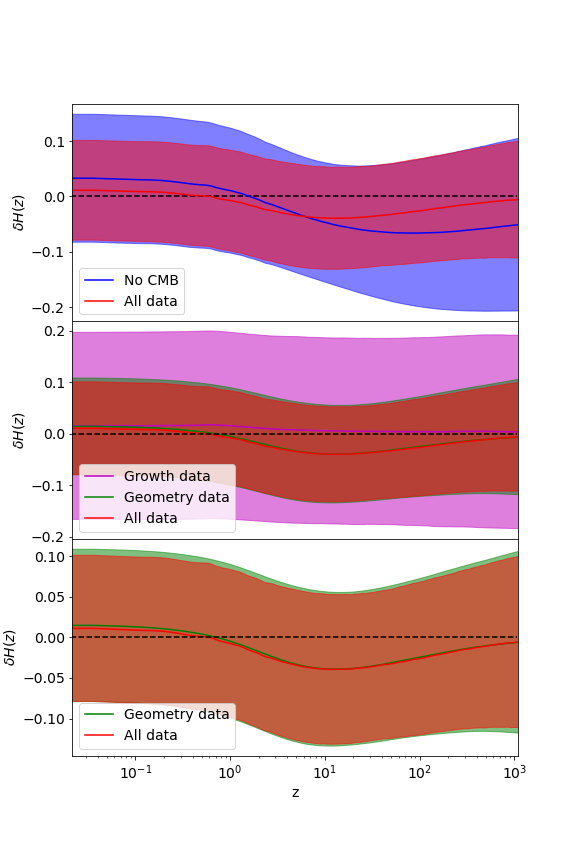} 
    \caption{$1\sigma$-constraints on $\delta H(z)$ broken down by type of data considered. Solid lines represent the mean of the GPs at each redshift. In red we display the constraints resulting from the analysis of all present data, in blue the effect of removing the CMB data set, in black the effect of fixing the $\Omega_m$ and $\sigma_8$ to their best-fit (BF) value; in magenta, the constraints resulting of only considering growth data; and in green, the constraints resulting of only considering geometry data.}
    \label{fig:dH_geo_vs_gro}
\end{figure}

Before presenting the results for the cosmological functions of interest; i.e. the expansion history $H(s)$ and linear growth rate $f \! \sigma_8$, we must first discuss the results on the core of our analysis, the GP on $\delta H(s)$. We found that our analysis of current data produces $\delta H(s)$ constraints compatible with $0$ at all redshifts at less than $1\sigma$ deviation. This means that the mean of the GP, $H_m(s) = A_0 \overline{H}(s)$, is capable of capturing the long range correlations of the data allowing the GP to fit local features. The recovered bounds on $\delta H$ are shown in Fig. \ref{fig:dH_geo_vs_gro}.

Moreover, we quantified how well different combinations of data can constrain $\delta H(s)$. In other words, we measured how strong the agreement of $\delta H(s)$ with zero is, and how it is affected by the data considered and the analysis choices. However, the fact that $\delta H(s)$ has a multivariate distribution means that there is not a unique figure of merit for how well data constrains it. In this work we focused in two numbers. On the one hand, we looked at the hyperparameter $\eta$ that constrains the prior distribution of values that $\delta H(s)$ can take. On the other hand, we computed the mean variance of the $\delta H(s)$ samples across redshift. In other words, for each parameter of $\delta H(s)$ (i.e. $\{\delta H(s_1),\,\delta H(s_2),\,...,\,\delta H(s_{200})\}$), we computed the variance of the \textsc{HMC} samples. Then, we took mean value of those variances, which is equivalent to averaging over redshift. We refer to this statistic as $\overline{\sigma}(\delta H(s))$. The motivation behind $\overline{\sigma}(\delta H(s))$ lies in the fact that it directly translates to average fractional constraints on $H(z)$ that can be readily interpreted. 

A summary of the impact of the different data sets and analysis choices on the distribution of $\eta$ and $\overline{\sigma}(\delta H(s))$ can be found in Tab. \ref{tab:variance}. For the combination of data sets employed in our fiducial analysis, we find $\eta = 0.113 \pm 0.075$ and an $\overline{\sigma}(\delta H(s)) = 0.094 \pm 0.004$ corresponding to an average $9.4 \pm 0.4\%$ constraint on $H(z)$ across redshift. Removing the CMB data point significantly worsened both constraints finding $\eta = 0.142 \pm 0.096$ and an average $12.5 \pm 1.6\%$ constraint on $H(z)$ across redshift. In order to better understand this effect we can look at the first panel in Fig. \ref{fig:dH_geo_vs_gro}. In this figure, we can observe how removing the CMB data from the analysis significantly widens the constraints of the GP, specially beyond $z>2.5$ (see Table \ref{tab:variance}). This is to be expected as the CMB data is the only point we have above $z=2.5$. Thus, the constraints from this integrated effect are expected to become dominant in the redshift range between $2.5 < z < 1100$. However, it is important to note that the CMB is not the only contributor to the $\delta H(s)$ constraints over this redshift range since $f \! \sigma_8$ data also constraints $\delta H(s)$ over its whole domain through its role in solving the Jeans equation. The associated constraint is however very weak.

For the purpose of studying the effect of different data types, we split the data points within the data sets employed in the fiducial analysis in two groups: ``geometry'' -- exclusively containing measurements of the expansion history -- and ``growth'' -- solely containing $f \! \sigma_8$ measurements. 

As we can see in the second panel of Fig. \ref{fig:dH_geo_vs_gro}, growth data only is much weaker  at constraining $\delta H(z)$. From Table \ref{tab:variance} we see that constraints from growth data alone on $\overline{\sigma}(\delta H(s))$ are approximately twice as wide as those resulting from analysing the entire data set. These constraints are consistent with the prior on the hyperparameter $\eta$. On the other hand, the constraining power of the geometry data is only slightly weaker than that of the entire data set. Hence the $\delta H(z)$ constraints are mostly dominated by the geometry data sets as one would expect. Nonetheless, the addition of growth data increases constraining power. This is shown explicitly in the last panel of Fig. \ref{fig:dH_geo_vs_gro}, which shows the results of using geometry data alone compared to those using the full data set. This recovers the expected behavior: more data increases the constraining power and the contours shrinks.

\begin{figure} 
\centering
    \includegraphics[width=\linewidth]{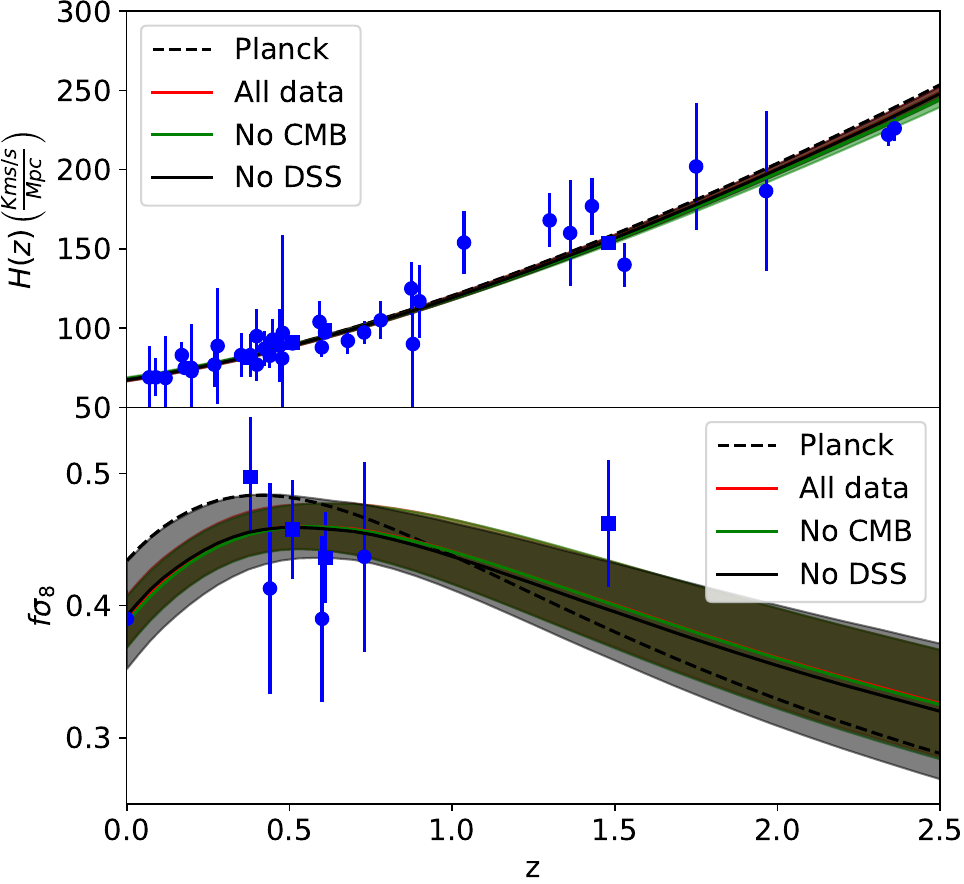} 
    \caption{$1\sigma$-constraints for the cosmological functions $H(z)$ and $f \! \sigma_{8}$ (top and bottom panel respectively) broken down by combination of data set. Solid lines represent the mean of the GPs at each redshift. The dashed black lines show the prediction for each cosmological function using our Planck 2018 fiducial cosmology (see Table~\ref{tab:fiducial}). In red we show the constraints resulting of the analysis of all present data; in green, the impact of removing the CMB data set; and in black, the impact of removing the DSS data set.}
    \label{fig:comp_data}
\end{figure} 

We now shift the focus of our discussion to the constraints we derive from $\delta H(s)$ for the expansion history itself, $H(z)$, and the linear growth of matter anisotropies, $f\!\sigma_8$. Comparing the constraints for both cosmological functions from our analysis of current data with the \textit{Planck} 2018 predictions, we find an overall good agreement, finding both functions to contain the \textit{Planck} 2018 predictions within their $2\sigma$ confidence contours. This can be seen in  Fig. \ref{fig:comp_data}. Nonetheless, two remarks can be made. First, we observed a greater than 1$\sigma$ preference for a lower $f\!\sigma_8$ between $0<z<0.75$, mostly driven by the DSS data point. However, the constraining power of current $f\!\sigma_8$ data is too weak to make a case for the presence of new physics. Second, our model-independent analysis finds the supernova absolute magnitude parameter to be $M=-19.43 \pm 0.026$, a constraint which is in $5 \sigma$ tension with the SHOES preferred value \citep{Efstathiou21}. However, this is not surprising given the known tension between the data sets used to inform the GP reconstruction.  

\begin{figure*} 
\hspace*{-0.7cm} 
\centering
    \includegraphics[width=\linewidth]{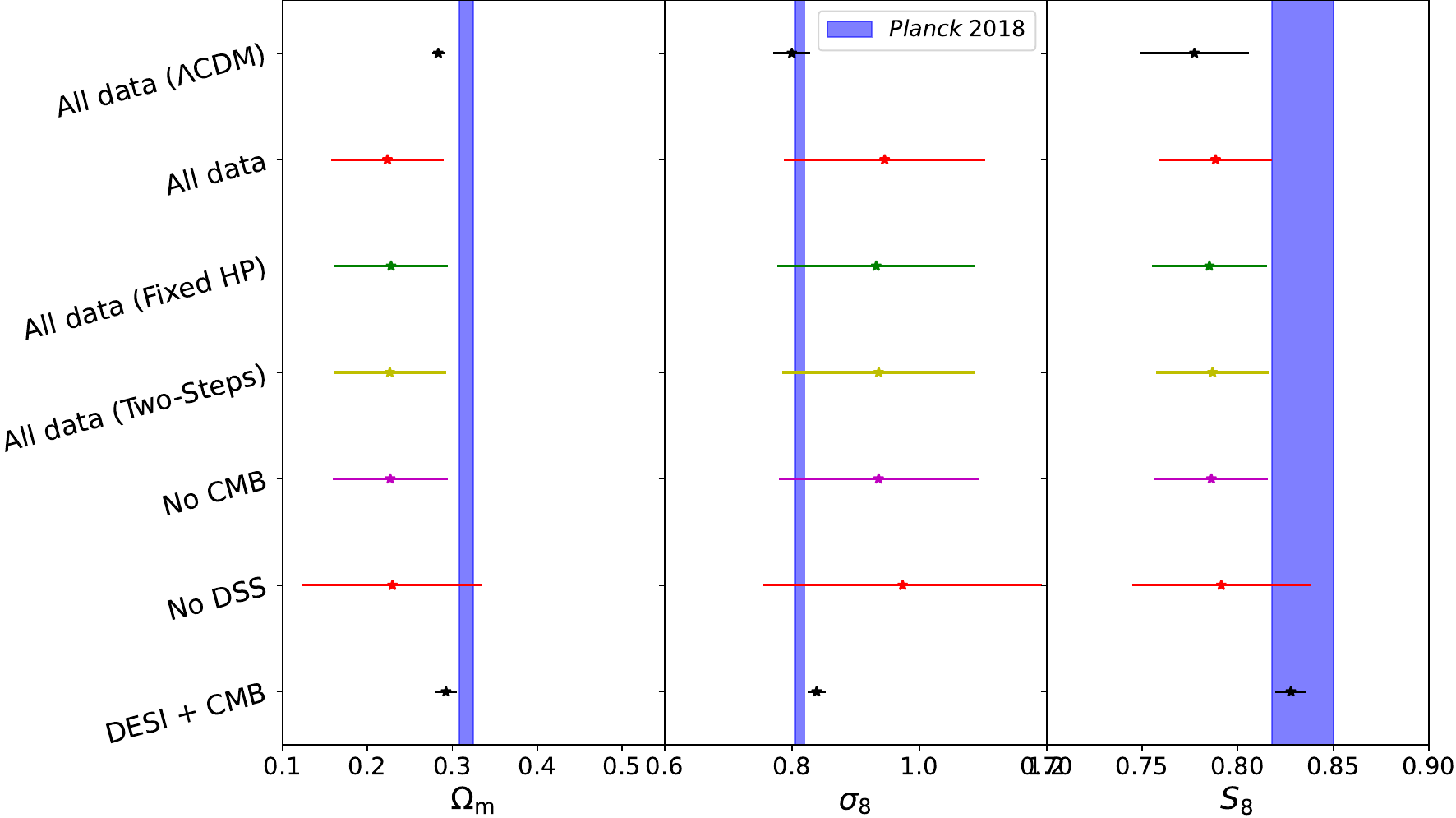} 
    \caption{Comparison between the different $1 \sigma$ constraints for the cosmological parameters $S_8$,  $\sigma_{8}$ and $\Omega_m$ (first, second and third panel respectively). In each panel the \textit{Planck} 2018 constraint is displayed in the form of a blue band for comparison.}
    \label{fig:1D_comp}
\end{figure*} 

\subsubsection{Cosmological parameters}
We now focus on the constraints on the specific cosmological parameters $\Omega_M$ and $\sigma_8$. We will also show constraints on the derived parameter $S_8 = \sigma_8 \sqrt{ \Omega_m / 0.3}$ and on the local value of the expansion rate, given in our case by $H_{0}\equiv A_0 \, \overline{H}_0 \, (1+ \delta H(s=0))$. A summary of the constraints obtained by the different analyses considered in this work can be found in Table \ref{tab:constraints} with a graphical illustration in Fig \ref{fig:1D_comp}. Full 2D contours for the respective sets of constraints, including hyperparameters, can be found in Appendix \ref{app:full_results}. We primarily focus our discussion on the cosmological parameter $\Omega_m$, the only remaining degree of freedom in Eqn. \ref{eq:growth}. 

\begin{table*} 
\caption{Constraints on the cosmological parameters $\Omega_m$, $\sigma_8$, $S_8$,  and $H_0$ (first to fourth columns respectively) for each of the different analyses (rows), as well as the \textit{Planck} 2018 constraints and $\Lambda$CDM analyses of current and DESI+CMB for reference (first to third row respectively).}
\begin{tabular}{|p{3cm}|p{1.75cm}|p{1.75cm}|p{1.75cm}|p{3cm}| }
 \hline
  Analysis & $\Omega_m$ & $\sigma_{8}$ & $S_8$ & $H_0 = \Bar{H_{0}} A_0 (1+\delta H(0))$ \\ 
    \hline
    
    \textit{Planck} 2018 &   $0.317 \pm 0.008$ &  $0.812 \pm 0.007$  & $0.834 \pm 0.016$ &  $67.27 \pm 0.006 $\\
    All data ($\Lambda$CDM) & $0.283 \pm 0.007$ & $0.8 \pm 0.029$ & $0.777 \pm 0.028$ & $68.601 \pm 0.775$ \\
    DESI+CMB ($\Lambda$CDM) & $0.316 \pm 0.006$ & $0.812 \pm 0.003$ & $0.834 \pm 0.008$ & $66.992 \pm 0.311$ \\
    \hline
    All data & $0.224 \pm 0.066$ & $0.946 \pm 0.158$ & $0.788 \pm 0.029$ & $67.715 \pm 0.93$ \\
    All data (Fixed HP) & $0.228 \pm 0.067$ & $0.932 \pm 0.155$ & $0.785 \pm 0.03$ & $68.113 \pm 1.061$ \\
    All data (Two-Steps) & $0.226 \pm 0.066$ & $0.936 \pm 0.151$ & $0.787 \pm 0.029$ & $68.144 \pm 0.718$ \\
    No CMB & $0.227 \pm 0.068$ & $0.936 \pm 0.156$ & $0.786 \pm 0.03$ & $67.94 \pm 1.034$ \\
    No DSS & $0.229 \pm 0.106$ & $0.974 \pm 0.218$ & $0.791 \pm 0.047$ & $67.766 \pm 0.96$ \\
    DESI $+$ CMB & $0.293 \pm 0.013$ & $0.839 \pm 0.014$ & $0.828 \pm 0.008$ & $66.788 \pm 0.371$ \\
\end{tabular}
\label{tab:constraints}
\end{table*}

To establish a benchmark against which to compare our method, we start by examining the constraints obtained assuming a $\Lambda$CDM model in which we vary $H_0$, $\Omega_m$ and $\sigma_8$. This allows us to quantify the impact of performing a model-independent analysis using GPs on the final constraining power. Looking at Fig. \ref{fig:1D_comp} and Table \ref{tab:constraints} we can observe that, assuming $\Lambda$CDM,  $\Omega_m = 0.283 \pm 0.007$. In other words, for the setup used in this work, we can constrain $\Omega_m $ to around $2\%$ precision if we undertake a model-dependent analysis with  $\Lambda$CDM. In this case, $\Omega_m$ receives information from both background and perturbations.

In turn, our fiducial model-independent analysis yields $\Omega_m = 0.224 \pm 0.066$, inflating the uncertainty by a factor of $\sim9$. Comparing this result with the \textit{Planck} 2018, $\Omega_m = 0.317 \pm  0.008$, our $\Omega_m$ constraint is lower but statistically compatible with the \textit{Planck} 2018 constraint and our $\Lambda$CDM analysis both at $1.5 \sigma$.

Looking at the relevance of the different individual data sets on the constraints, we observe an excellent agreement between all the different combinations considered (see Fig. \ref{fig:1D_comp} and Tab. \ref{tab:constraints}).  Removing the DSS data, one of the most precise $f \! \sigma_8$ measurements, significantly worsens the $\Omega_m$ constraints by nearly $60 \%$. On the other hand, removing the CMB data set resulted in nearly identical constraints for the cosmological parameters. This is due to the fact that, in the presence of other geometry data to inform the expansion history, $\Omega_m$ constraints become dominated by growth data through the relationship between both. Thus, it is fair to ask what the impact of completely removing any type of geometry data from the analysis is. Analytically, we can see from Eqn. \ref{eq:jeans} that in the presence of measurements of the linear growth rate and an arbitrary value of $\Omega_m$ one can always find a expansion history that solves the differential equation. This degeneracy between the expansion history and $\Omega_m$ in Eqn. \ref{eq:jeans}, prevents our methodology from obtaining any meaningful constraints on $\Omega_m$ in the absence of Geometry data that is not completely dominated by our choice of GP mean and hyperparameter priors.

Looking at the parameters $\sigma_8$ and $S_8$ in more detail, we found compatible constraints with the \textit{Planck} 2018 cosmology up to 1$\sigma$. However, it is worth mentioning that our results show a mild tendency towards higher $\sigma_8$ values which, combined with the tendency towards lower  $\Omega_m$ values, results in lower $S_8$ values. This is consistent with the underprediction of $f \! \sigma_8$ between $0<z<0.75$ we discussed in the previous section. A lower value of $S_8$ would also be consistent with the most recent measurements by large scale structure experiments \citep{K1K, DESY3, Garcia-Garcia21, 2021arXiv211109898W, 2021arXiv210503421K}, the origin of which could lie on a lower value of $\Omega_m$ \citep{2021MNRAS.501.1481H}.

\subsection{Forecasts}
\begin{figure} 
\centering
    \includegraphics[width=\linewidth]{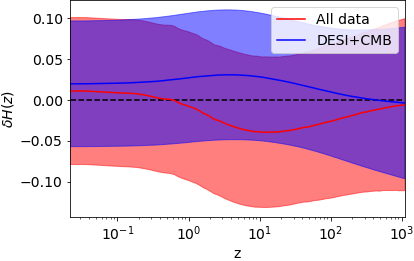} 
    \caption{Forecast $1\sigma$-constraints on $\delta H(z)$ from DESI (blue) and using current data (red). Solid lines represent the mean of the GPs at each redshift.}
    \label{fig:dH_forecast}
\end{figure}

Our model-independent analysis leads to far weaker constraints than assuming the $\Lambda$CDM model. Being data-driven, the performance of the method used here may improve significantly with the advent of next-generation surveys with significantly tighter uncertainties. To quantify this, we repeated our fiducial analysis pipeline on mock data generated based on the forecasted errors for the DESI mission \citep{Font-Ribera14} in combination with the \textit{Planck} measurement of $\theta_*$. The results can be found in Fig. \ref{fig:dH_forecast}. In this figure we can observe a $10\%$ improvement between the expected $\delta H(z)$ constraints from DESI with respect to those of current data. The corresponding $\eta$ and $\overline{\sigma}(\delta H(s))$ constraints (shown in Tab. \ref{tab:variance}) improve by approximately $15\%$: $\eta = 0.085 \pm 0.074$ and $\overline{\sigma}(\delta H(s)) = 0.080 \pm 0.004$. Looking at the cosmological parameters,  DESI \citep{Font-Ribera14} in combination with the CMB data set results in nearly 5 times tighter $\Omega_m$ constraints and 10 times better constraints on $\sigma_8$.

While these constraints are still significantly worse than the \textit{Planck} 2018 results, they are comparable if not better than using the $\Lambda$CDM model to analyze the currently available data used in this work. For example, while GP constraints on $\Omega_m$ from DESI are two times wider than those currently found assuming a $\Lambda$CDM model, the constraints on $\sigma_8$ and $S_8$ are 2.3 and 3.6 times tighter respectively. 

If we, instead, compare how a model-independent analysis of DESI data using this methodology will pitch against a $\Lambda$CDM analysis, we find that the gap between the model-independent and the model-dependent constraints shrinks. This can be see in the fact that, with current data, the $\Lambda$CDM constraint on $\Omega_m$ is $9.2$ times tighter than the one obtained with our methodology. However, in our DESI forecast, it is only twice as good. Similarly, the improvement for the parameter $\sigma_8$ reduces from five times tighter constraints to about four times. Our understanding is that, as better data allow us to better reconstruct $H(z)$, the difference between the reconstructed and model-dependent $H(z)$ shrinks, as long as the assumed model fits the data well. As a consequence, the constraints on the other parameters ($\Omega_m$, $\sigma_8$ and $S_8$) become more similar. Thus, we expect that in the future, as the quality of the data keeps improving, cosmological constraints from model-independent methodologies, such the one proposed in this work, will rival model-dependent ones.

\section{Conclusions} 
\label{Sect: Conclusions}
 In this work we have developed a method to obtain  constraints on $H(z)$ and $\Omega_m$ purely based on the relationship between the expansion history and the linear growth rate. In order to do so, we employ a Gaussian process to model the evolution of the expansion history from present time to recombination.

From this expansion history, we have derived predictions for the comoving distance and the linear growth rate $f \! \sigma_8$ making use of the physical relationships between the three quantities. Constraints for $\Lambda$CDM parameters were obtained by simultaneously fitting a suite of the latest measurements of these three cosmological functions. The data combination used for our fiducial analysis consisted of cosmic chronometers, the Pantheon supernova catalog, BAO and RSD data, peculiar velocity data from supernovae, and the position of the first acoustic peak in the CMB power spectrum. Moreover, we also obtained forecast constraints on these cosmological functions from the future DESI data.

Current data can constrain the $H(z)$ Gaussian process at an average $9.4\%$ throughout all redshifts. These constraints are compatible up to $1 \sigma$ with the best-fit \textit{Planck} 2018 expansion history across $0 < z < 1100$. Our constraints on the expansion rate $f \! \sigma_8$ lie below the corresponding \textit{Planck} prediction by less than 2-$\sigma$ in the range $0 < z < 0.75$ (a result driven by the Democratic Supernova Sample data).

Translating the Gaussian process constraints into constraints of cosmological parameters, we find a model-independent measurement of $\Omega_m = 0.224 \pm 0.066$. This result is lower than, but statically compatible with, the Planck 2018 \citep{Planck} cosmology. We also find $S_8 = 0.788 \pm 0.029$, an intermediate value, statistically compatible with both the Planck 2018 cosmology \citep{Planck} and recent local measurements from weak lensing and galaxy  clustering\citep{DESY3, K1K, Garcia-Garcia21, 2021arXiv210503421K, 2021arXiv211109898W}.

The forecast analysis performed using the methodology of this work predicts that combining the DESI measurements with the CMB BAO data used in this work will yield $15\%$ tighter constraints on $H(z)$ across redshift as well as five times tighter constraints on $\Omega_m$. While this constraints are still significantly looser than the model-dependent cosmological constraints of \textit{Planck} 2018, we show that a model-independent analysis of DESI plus CMB BAO data would be as powerful as a $\Lambda$CDM analysis of current data. Indeed, we find that model-independent constraints using this methodology would achieve four times tighter $S_8$ constraints. Thus, in the future it will be possible to weigh in on the ongoing $S_8$ tension making use of model-independent methods. Moreover, we also show that, as the quality of the data increases as we go into the future, the gap between the constraining power of model-independent and model-dependent constraints will significantly shrink. 

Future implementations of this methodology could explore several possible extensions. On the one hand, the types of data used in this work are greatly limited by lack of differentiable tools to obtain theoretical predictions for observables with greater constraining power. The development of tools such as differentiable Boltzmann codes, differentiable emulators of the non-linear matter power spectrum, or differentiable Limber integrators would allow us to fit the power spectrum data directly. These developments would enable model-agnostic analyses similar to that of \citet{Garcia-Garcia21} but with a greatly reduced number of assumptions and a more reliable measure of uncertainty in their results. On the other hand, it would also be possible to explore the use of Gaussian processes to constrain general forms of modified gravity, generalizing works such as \citet{2017PhRvD..96h3509R, 2019PhRvD..99b3512E, 2021PhRvD.103j3530P, 2021arXiv210712990R, 2021arXiv210712992P}, and study how these theories can be informed by the relationship between background and perturbations. Gaussian processes are an exceptional tool to constrain modified gravity since they don't require assuming a particular parametrisation of such deviations. A comprehensive list of different departures from $\Lambda$CDM that could be explored with a similar methodology to the one presented in this work can be found in \citet{Baker14}. Alternatively, one could consider comparing the performance of Gaussian processes against other popular tools for non-parametric cosmology such as genetic algorithms \citep{ga, Nesseris11} or neuronal networks \citep{ANN}. The convenience of using Gaussian processes lies in that they naturally provide easily interpretable information on both the reconstructed function and its posterior uncertainties. 

\section*{Acknowledgements}
\textit{Author contributions}: All authors contributed to the development and writing of
this paper. We acknowledge support from the Beecroft Trust. CGG and PGF acknowledge funding from the European Research Council (ERC) under the European Unions Horizon 2020 research and innovation programme (grant agreement No 693024). DA acknowledges support from the Science and Technology Facilities Council through an Ernest Rutherford Fellowship, grant reference ST/P004474. JRZ is supported by an STFC doctoral studentship.

The analysis made use of the software tools
\textsc{SciPy} \citep{scipy}, \textsc{NumPy} \citep{np}, \textsc{Matplotlib} \citep{np},
\textsc{CLASS} \citep{Class}, GetDist \citep{getDist}.

Based on observations obtained with \textit{Planck} (\url{http://www.esa.int/Planck}), an ESA science mission with instruments and contributions directly funded by ESA Member States, NASA, and Canada.

Funding for the Sloan Digital Sky Survey IV has been provided by the  Alfred P. Sloan Foundation, the U.S. Department of Energy Office of Science, and the Participating Institutions. SDSS-IV acknowledges support and  resources from the Center for High Performance Computing  at the University of Utah. The SDSS website can be found at \url{www.sdss.org}.
SDSS-IV is managed by the  Astrophysical Research Consortium for the Participating Institutions of the SDSS Collaboration including  the Brazilian Participation Group,  the Carnegie Institution for Science,  Carnegie Mellon University, Center for Astrophysics | Harvard \& Smithsonian, the Chilean Participation, the French Participation Group,  Instituto de Astrof\'isica de Canarias, The Johns Hopkins  University, Kavli Institute for the Physics and Mathematics of the Universe (IPMU) / University of Tokyo, the Korean Participation Group, Lawrence Berkeley National Laboratory, Leibniz Institut f\"ur Astrophysik Potsdam (AIP),  Max-Planck-Institut f\"ur Astronomie (MPIA Heidelberg), Max-Planck-Institut f\"ur  Astrophysik (MPA Garching), Max-Planck-Institut f\"ur Extraterrestrische Physik (MPE), National Astronomical Observatories of China, New Mexico State University, New York University, University of Notre Dame, Observat\'ario Nacional / MCTI, The Ohio State University, Pennsylvania State University, Shanghai Astronomical Observatory, United Kingdom Participation Group, Universidad Nacional Aut\'onoma de M\'exico, University of Arizona, University of Colorado Boulder, University of Oxford, University of Portsmouth, University of Utah, University of Virginia, University of Washington, University of Wisconsin, Vanderbilt University, and Yale University.

\section*{Data Availability}
The code developed for this work as well as the data sets used are available upon request. 

%%%%%%%%%%%%%%%%%%%% REFERENCES %%%%%%%%%%%%%%%%%%

% The best way to enter references is to use BibTeX:

\bibliographystyle{mnras}
\bibliography{GP}

%%%%%%%%%%%%%%%%%%%%%%%%%%%%%%%%%%%%%%%%%%%%%%%%%%

\appendix

\section{Full results}
\label{app:full_results}
\begin{figure*}
\centering
    \includegraphics[width=\linewidth]{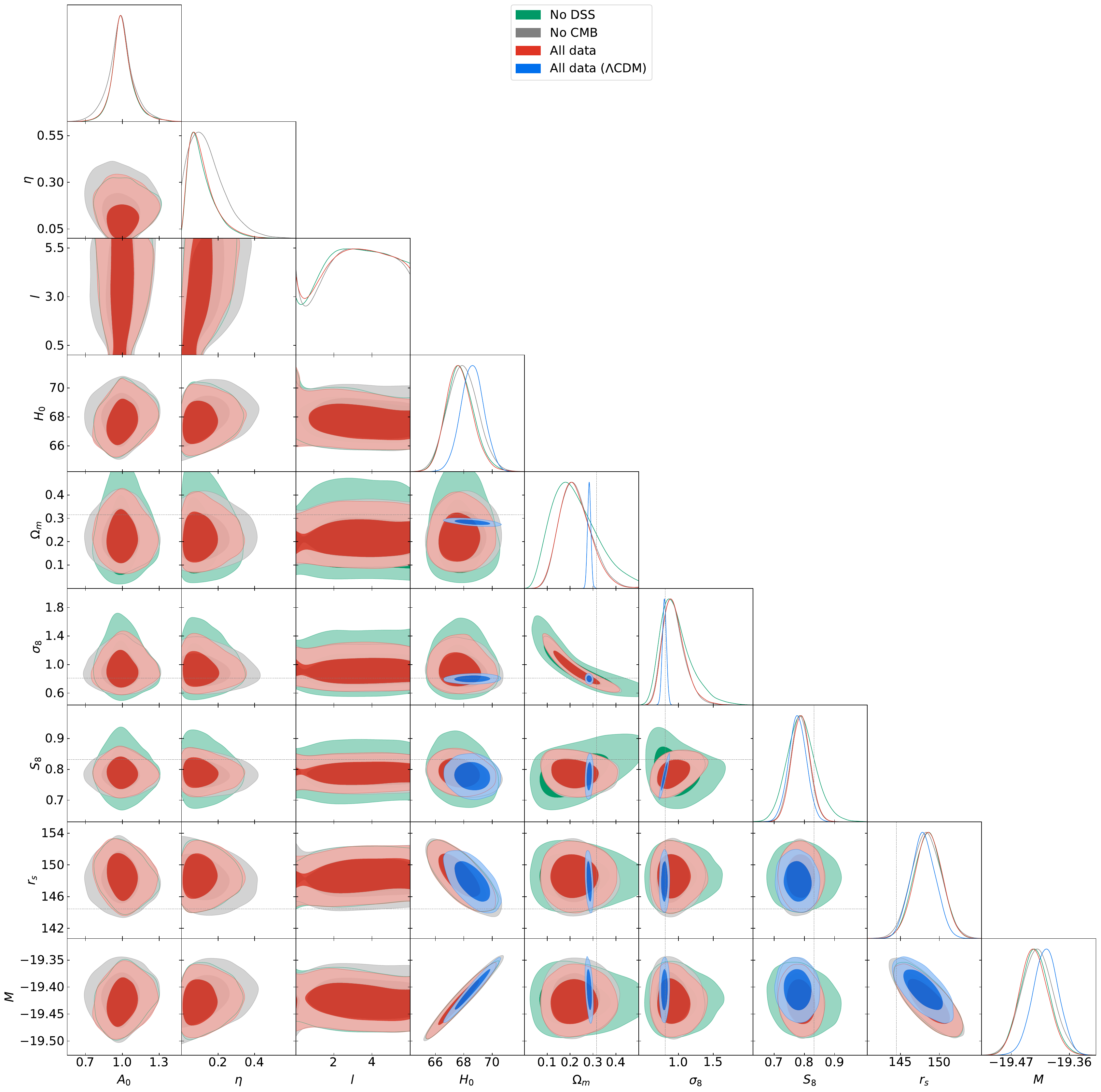} 
    \caption{ Comparison of the full posterior distributions of the analysis of all the considered data (red), the analysis of all data except for the DSS data set (green), the analysis of all data except for the CMB data set (grey) and the $\Lambda$CDM analysis of the data (blue). The parameters displayed are the GP hyperparameters $\eta$ and $l$, the cosmological parameters $H(z=0)$, $\Omega_m$, $\sigma_8$ and $S_8$ in this order and the nuisance parameters $r_s$ and $M$. Central panels show the 2D histograms of the different parameter combinations while diagonal panels show 1D histograms.}
    \label{fig:triangle_data}
\end{figure*}

\begin{figure*} 
\centering
    \includegraphics[width=\linewidth]{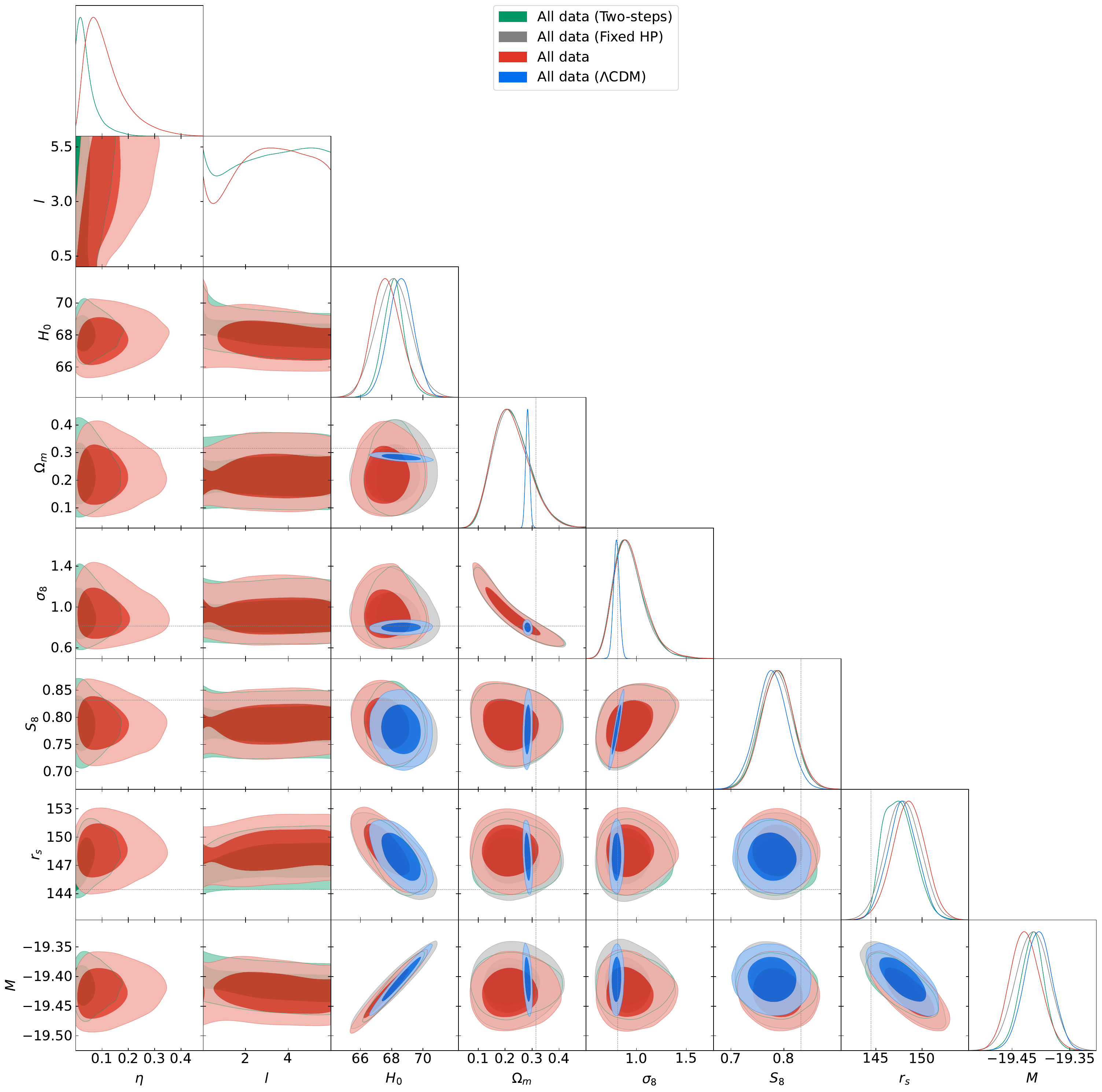} 
    \caption{ Comparison of the full posterior distributions of the of the standard analysis of all considered data (red), the Two-Steps analysis constraints (green), the fixed hyperparamters analysis (grey) and the $\Lambda$CDM analysis of the data (blue). The parameters displayed are the GP hyperparameters $\eta$ and $l$, the cosmological parameters $H(z=0)$, $\Omega_m$, $\sigma_8$ and $S_8$ in this order and the nuisance parameters $r_s$ and $M$. Central panels show the 2D histograms of the different parameter combinations while diagonal panels show 1D histograms.}
    \label{fig:triangle_tests}
\end{figure*}

\begin{figure*} 
\centering
    \includegraphics[width=\linewidth]{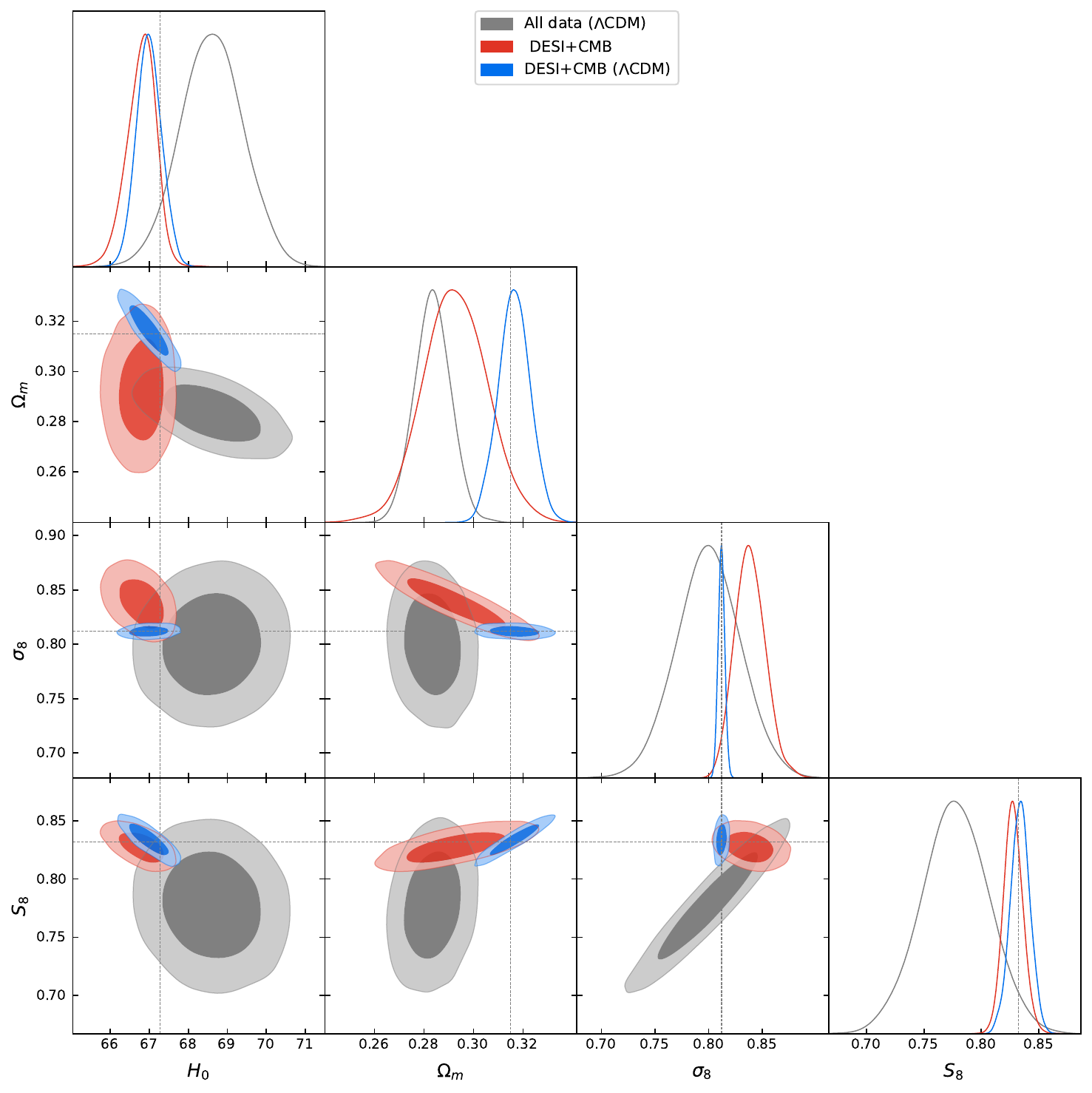} 
    \caption{ Comparison of the full posterior distributions between the DESI+CMB analysis (blue) and the $\Lambda$CDM analysis of present data (red). The parameters displayed are the cosmologoical parameters $H(z=0)$, $\Omega_m$, $\Omega_b$, $\sigma_8$ and $S_8$ in this order. Central panels show the 2D histograms of the different parameter combinations while diagonal panels show 1D histograms.}
    \label{fig:triangle_forecast}
\end{figure*}

In this appendix we present the full set of posterior distributions for all cosmological parameters and for each of the combinations of data sets and analysis choices studied here. In Fig. \ref{fig:triangle_data} we show the marginalised posteriors resulting of the different combinations of data sets. In Fig. \ref{fig:triangle_tests} we show the marginalised posteriors resulting of the different analysis choices for the entire combination of data sets juxtaposed.  In Fig. \ref{fig:triangle_forecast} we show the marginalised posteriors using the synthetic DESI dataset.

\section{Alternative analyses}
\label{app: Alternative analyses}

\begin{figure} 
\centering
    \includegraphics[width=\linewidth]{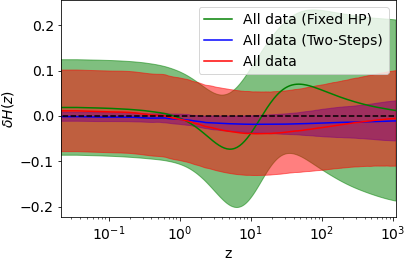} 
    \caption{ Comparison of the constraints for the cosmological functions $\delta H(z)$ between the standard analysis (red), the fixed hpyerparameters analysis (green, Fixed HP) and the Two-Steps analysis (blue) of the all considered data. }
    \label{fig:dH_tests}
\end{figure}

In this appendix we present the impact on the constraints on $\delta H(z)$ of the different analysis choices enumerated at the end of Sect. \ref{Sect: Methodology}. Specifically, we studied the effect of fixing the value of the hyperparameters $\eta$ and $l$ to the values listed in Table \ref{tab:prior} (labeled 'Fixed HP') and the effect of performing a two-step analysis (labeled ``Two-Steps''). In Fig. \ref{fig:dH_tests}, we can observe how fixing the hyperparameters  drastically changes the shape of $\delta H(z)$. On the one hand, fixing the value of $\eta$ prevents the low redshift data from constraining its value. Meaning that in the absence of data (i.e. for $z>2.5$), the GP amplitude is constant and completely dominated by the prior as opposed to being inferred from the low redshift data. On the other hand, fixing $l=1.0$ imposes a particular scale of correlation on the expansion history. 

Performing a two-step analysis produces qualitatively similar results to simultaneously re-scaling the GP mean with the parameter $A_0$. However, the two-step analysis produces significantly tighter constraints on $\delta H(z)$ due to the breaking of the degeneracies between $A_0$ and $\delta H(s)$. This can be observed in the two-step analysis recovering a three times smaller value of the hyperparameter $\eta$ and a four times smaller value the $\overline{\sigma}(\delta H(s))$ statistic with respect to simultaneously re-scaling the mean. This is due to the fact that the parameter $A_0$, used to re-scale the mean, is partially degenerate with the amplitude of $\delta H(z)$ and thus with $\eta$, introducing a greater degree of uncertainty.

\begin{figure} 
\centering
    \includegraphics[width=\linewidth]{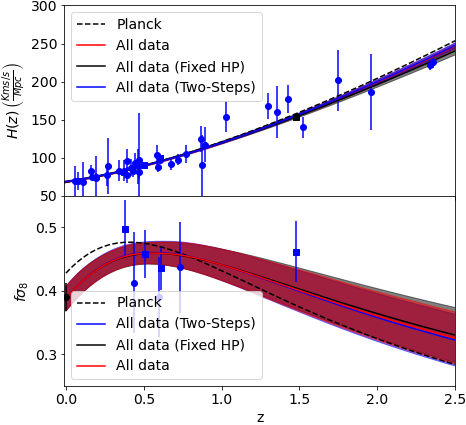} 
    \caption{ Comparison of the constraints for the cosmological functions $H(z)$ and  $f \! \sigma_{8}$ (top and bottom panel respectively) between the standard analysis (red), the Two-Steps analysis (blue) and the fixed hyperparameters analysis (grey) of all considered data. Solid lines represent the mean of the GPs at each redshift. Contours represent the associated $1\sigma$ confidence regions. The dashed black lines show the \textit{Planck} 2018 prediction for each cosmological function.}
    \label{fig:comp_tests}
\end{figure}

If instead of looking at $\delta H(z)$ we focus on the cosmological functions $H(z)$ and $f\!\sigma_8$, we find that the predictions of the fiducial and alternative analyses for the cosmological functions are statistically compatible up to 1$\sigma$ for all redshifts. This can be observed in Fig. \ref{fig:comp_tests}. This means that despite the fact that the analysis choices can have a severe impact in the shape of $\delta H(z)$, once $\delta H(z)$ is converted into the cosmological functions (i.e. the quantities actually being matched to the data), the standard and alternative analyses converge to similar predictions. 

Reassuringly the constraints for the cosmological parameters $\Omega_m$, $\sigma_8$ and $S_8$ are virtually unaffected by either fixing the hyperparameters or choosing a Two-Steps analysis over simultaneously fitting the mean of the GP. This can be seen in both Fig. \ref{fig:1D_comp}. It would be interesting to understand up to what degree of uncertainty on $H(z)$ one recovers the same constraints for these parameters. We leave this for future work.

Preliminary analyses showed that, if instead of constraining the cosmological parameters of the mean of the GP in a separate step, they were sampled simultaneously with the GP in a single step, one obtains a similar level of uncertainty to that introduced by the $A_0$ parameter. Nonetheless, $A_0$ has the added benefit of not duplicating parameters. For this reason we favour the scaling with $A_0$ as our main methodology.

\begin{figure} 
\centering
    \includegraphics[width=\linewidth]{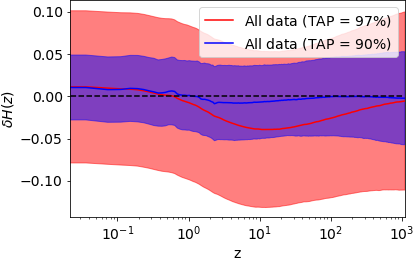} 
    \caption{ Comparison between the $\delta H(z)$ constraints obtained from two different \textsc{HMC} runs, one in which the target acceptance probability was set to $90 \%$ and one in which it was set to $97 \%$. In both cases all present data was considered. Solid lines represent the mean of the GPs at each redshift. Contours represent the associated $1\sigma$ confidence regions. }
    \label{fig:90_vs_97}
\end{figure}

\section{Validity of cosmological inference}
\label{app:validation}
In order to verify that the cosmological constraints obtained by this methodology are not biased by the choice of mean for the GP, we performed a blind analysis of mock data generated from five known cosmologies different from \textit{Planck} 2018 named Cosmo1-5. The mock data was generated following the errors and redshift of current measurements to ensure the validity of the methodology when applied to real data. We display the chosen cosmological parameter values in Tab. \ref{tab:challenge}. Cosmo1 corresponds to a \textit{Planck} 2018 cosmology. Cosmo2 posses a slightly higher value of $\Omega_m$ with respect the previous. Cosmo3 has a equation of state parameter $w_0 > -1$. Cosmo4 has significantly larger value of $H_0$ than  \textit{Planck} 2018. Finally, Cosmo5 has a significantly larger value of $\Omega_m$ than  \textit{Planck} 2018. 

\begin{table*} 
\caption{ Values of the cosmological parameters used to define the alternative cosmologies used to generate the mock data for the blind fold analysis.}
\centering
\begin{tabular}{ |p{1.5 cm}|p{1.5cm} |p{1.5cm} |p{1.5cm} |p{1.5cm} |p{1.5cm}}
 \hline
  Parameter & Cosmo1 & Cosmo2 & Cosmo3 & Cosmo4 & Cosmo5   \\
    \hline
    $\Omega_{\rm{m}}$   & 0.316 & 0.356 & 0.316  & 0.316 & 0.474 \\
    $\Omega_b$    & 0.050 & 0.056 & 0.050 & 0.050 & 0.070 \\
    $\Omega_{\Lambda}$ & 0.683 & 0.644 & 0.683 & 0.683 & 0.526\\
    %$\Omega_r$ &  $9.245 10^{-5}$ & $9.245 10^{-5}$ & $9.245 10^{-5}$ & $9.245 10^{-5}$ & $9.245 10^{-5}$ \\
    $H_0$ & 67.27 & 67.27 & 67.27 & 74 & 67.27\\
    %$n_{\rm{s}}$  & 0.9649 & 0.9649 & 0.9649 & 0.9649 & 0.9649  \\
    $\sigma_8$  & 0.811 & 0.811 & 0.811 & 0.811 & 0.811 \\
    $w_0$  & -1 & -1 & -0.9 & -1 & -1\\
\end{tabular}
\centering
\label{tab:challenge}
\end{table*}

\begin{figure*} 
\centering
    \includegraphics[width=\linewidth]{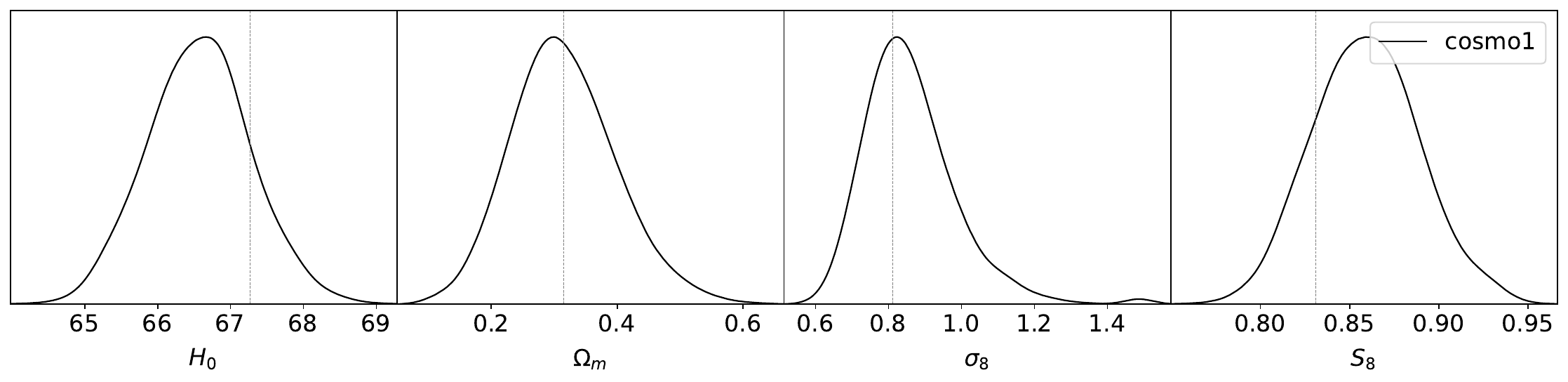} 
    \caption{ 1D Posterior distributions for the  cosmological parameters $H(z=0)$, $\Omega_m$, $\sigma_8$ and $S_8$ resulting from the blind analysis of one random realization of mock data generate from Cosmo1.  Dotted lines shows the value of the cosmological parameters used to generate the random realization of the data. Central panels show the 2D histograms of the different parameter combinations while diagonal panels show 1D histograms. Darker contours cover the $1\sigma$ CL region, whereas lighter ones, cover the $2\sigma$ region.}
    \label{fig:cosmo1}
\end{figure*}

\begin{figure*} 
\centering
    \includegraphics[width=\linewidth]{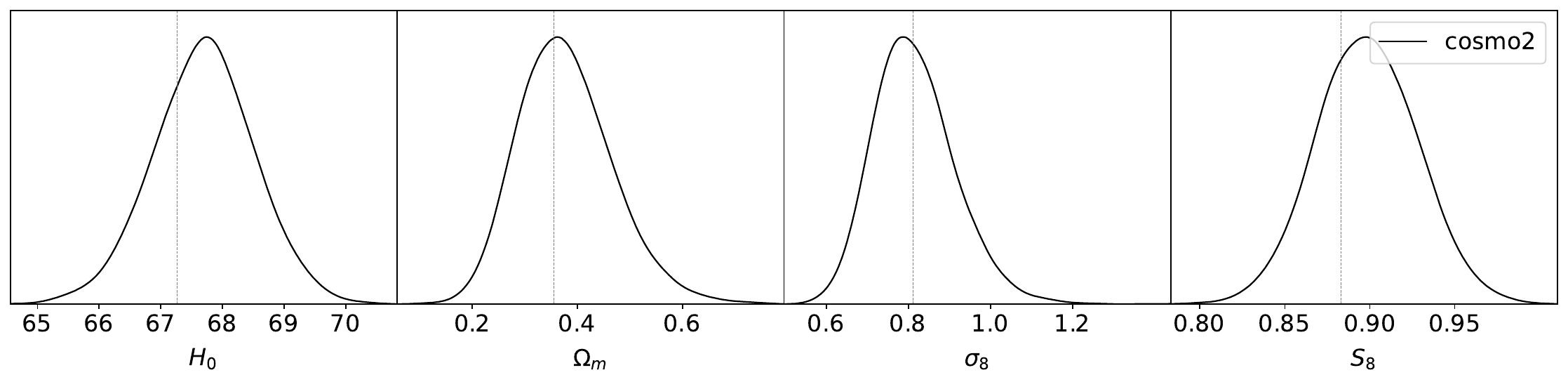} 
    \caption{ 1D Posterior distributions for the  cosmological parameters $H(z=0)$, $\Omega_m$, $\sigma_8$ and $S_8$ resulting from the blind analysis of one random realization of mock data generate from Cosmo2. Dotted lines shows the value of the cosmological parameters used to generate the random realization of the data. Central panels show the 2D histograms of the different parameter combinations while diagonal panels show 1D histograms.Darker contours cover the $1\sigma$ CL region, whereas lighter ones, cover the $2\sigma$ region.}
    \label{fig:cosmo2}
\end{figure*}

\begin{figure*} 
\centering
    \includegraphics[width=\linewidth]{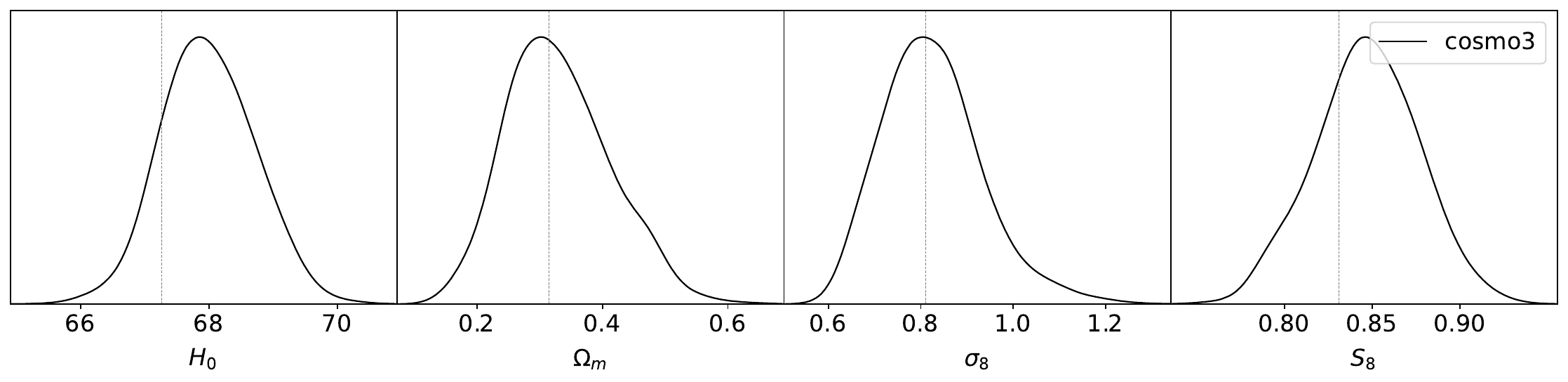} 
    \caption{ 1D Posterior distributions for the  cosmological parameters $H(z=0)$, $\Omega_m$, $\sigma_8$ and $S_8$ resulting from the blind analysis of one random realization of mock data generate from Cosmo3. Dotted lines shows the value of the cosmological parameters used to generate the random realization of the data. Central panels show the 2D histograms of the different parameter combinations while diagonal panels show 1D histograms.Darker contours cover the $1\sigma$ CL region, whereas lighter ones, cover the $2\sigma$ region.}
    \label{fig:cosmo3}
\end{figure*}

\begin{figure*} 
\centering
    \includegraphics[width=\linewidth]{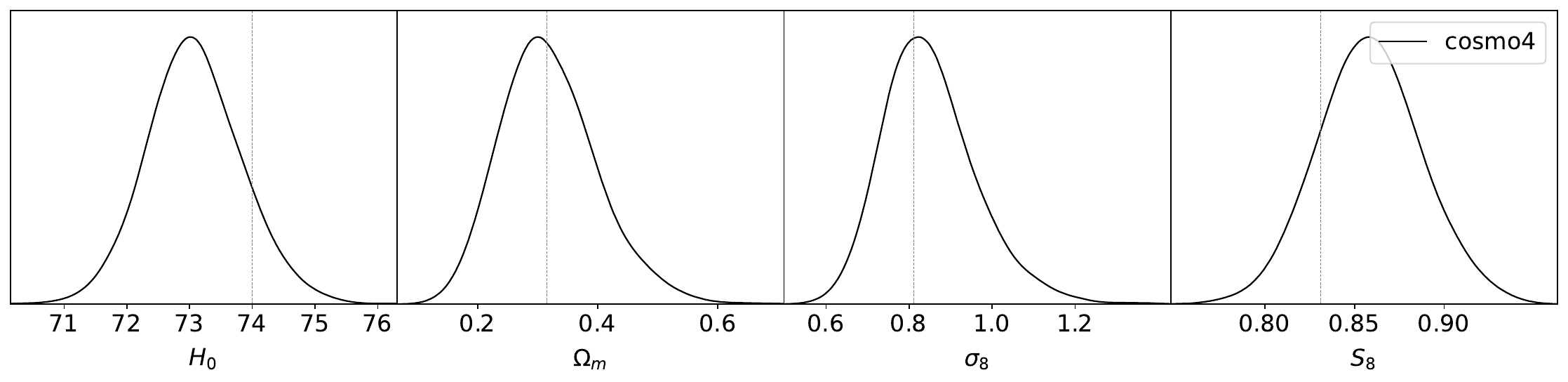} 
    \caption{ 1D Posterior distributions for the  cosmological parameters $H(z=0)$, $\Omega_m$, $\sigma_8$ and $S_8$ resulting from the blind analysis of one random realization of mock data generate from Cosmo4. Dotted lines shows the value of the cosmological parameters used to generate the random realization of the data. Central panels show the 2D histograms of the different parameter combinations while diagonal panels show 1D histograms. Darker contours cover the $1\sigma$ CL region, whereas lighter ones, cover the $2\sigma$ region.}
    \label{fig:cosmo4}
\end{figure*}

\begin{figure*} 
\centering
    \includegraphics[width=\linewidth]{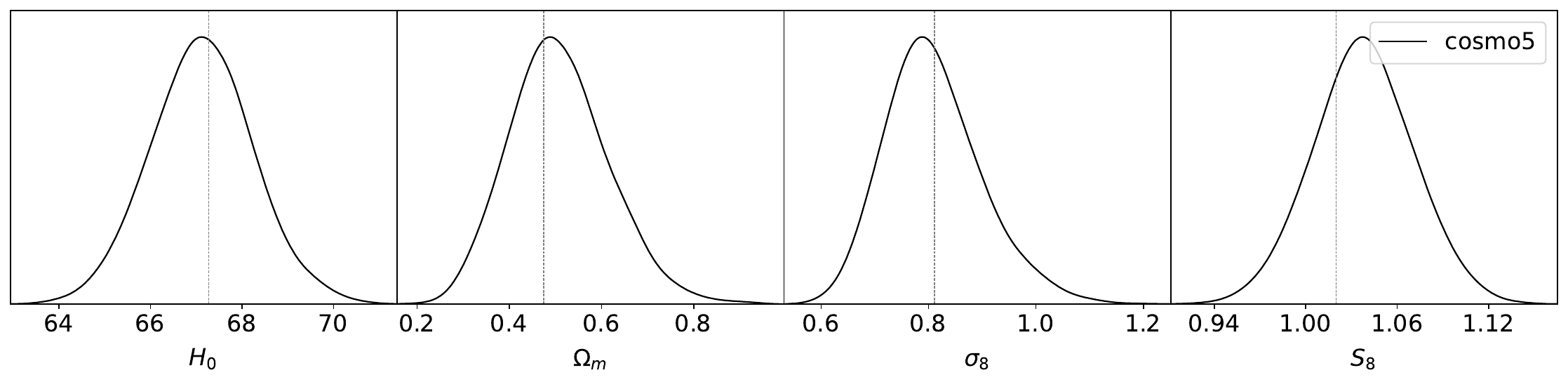} 
    \caption{ 1D Posterior distributions for the  cosmological parameters $H(z=0)$, $\Omega_m$, $\sigma_8$ and $S_8$ resulting from the blind analysis of one random realization of mock data generate from Cosmo5. Dotted lines shows the value of the cosmological parameters used to generate the random realization of the data. Central panels show the 2D histograms of the different parameter combinations while diagonal panels show 1D histograms. Darker contours cover the $1\sigma$ CL region, whereas lighter ones, cover the $2\sigma$ region.}
    \label{fig:cosmo5}
\end{figure*}

Figs \ref{fig:cosmo1}, \ref{fig:cosmo2}, \ref{fig:cosmo3}, \ref{fig:cosmo4} and \ref{fig:cosmo5} show the 1D posterior distributions obtained from analyzing one random realization of mock data from each of the cosmologies considered. Our results show that for all five scenarios our methodology returns cosmological constraints statistically compatible with the set of cosmological parameters used to generate the respective data sets. This can be observed in the fact that for all cases the 2 $\sigma$ contours of each cosmological parameter contain value of the underlying cosmology. 

However, previous iterations of our methodology where the value of the parameter $A_0$ was kept fixed at $A_0=1$ showed significant biases in their cosmological parameter inference with respect the input cosmological parameters. This biases manifested specially strongly in the constraints for the parameter $H_0$. This stresses the need of choosing a mean that does not systematically underpredict or overpredict the data being fitted with GP, specially if there are sections in the GP domain in which data is not present. In this range, the GP returns to its prior distribution, and thus a biased mean results on biased cosmological parameters and functions.

\section{Using HMC}
\label{app: Using HMC}

In this work we employed the \textsc{No U-Turn} sampler (\textsc{NUTS}), a self-tuning variant of the \textsc{Hamiltonian Monte Carlo} (\textsc{HMC}) sampler, implemented  in \textsc{Pymc3}. In this implementation, chains are divided in two stages: a tuning phase and an exploration phase. During the tuning phase, the \textsc{NUTS} algorithm tunes the so called "mass matrix". The mass matrix is the covariance matrix of the conjugate momentum parameters and  plays a role in how effectively the different energy values of the Hamiltonian are explored. Therefore, allowing for a sufficiently long tuning phase results in a well calibrated mass matrix which then translates into an efficient exploration stage of the chain.

However, even when equipped with a well informed mass matrix, HMC can be turned unreliable and inefficient if one overlooks the effectiveness of the numerical methods being used to resolved the Hamiltonian trajectories. Thus, a poor configuration of said methods can lead to the numerical solvers being unable to resolve the trajectories. These instances are known as divergences. Divergences have the undesired effects of a) making the sampling process inefficient and b) biasing the resulting posteriors. On the one hand, divergences make the sampling process inefficient as their fraction of the total number of steps needs to be rejected. On the other hand, divergences break the geometry ergodicity of sampler (i.e. its ability to explore the whole parameter space) such that stationary distribution of the Markov chains is no longer guaranteed to be target distribution. Divergences tend to occur in those areas of the likelihood function with high curvature (i.e. the slopes of the distributions). Thus, divergences can lead to artificial constraining power by virtue of the majority of the samples clustering at the regions of low likelihood curvature (i.e. the top of posterior distribution and the tails), creating a fake abrupt drop of the probability density outside the vicinity of the area of maximum likelihood. The \textsc{NUTS} implementation of \textsc{Pymc3}, allows to control such aspects by what it is referred to as the ``target acceptance probability'' (the desired probability of the next step of the Monte Carlo chain being accepted). \textsc{PyMC3} tunes the step size of the numerical methods during the tuning phase of the algorithm such that the desired target acceptance probability is obtained.
A target acceptance probability of one is thus equivalent to an infinitesimally small step size that perfectly resolves the Hamiltonian trajectories and for which no divergences occur.  However it is easy to see that such a small step size would result in an extremely slow exploration of the posterior. Therefore, it is vital to strike a balance between reliability and speed. In this work, we find that, for a our likelihood, a target acceptance probability of $97\%$ results in a both efficient and reliable exploration of the posterior distribution. 

In Fig. \ref{fig:90_vs_97} we illustrate the severity of the bias that such divergences can introduce in the apparent constraining power of the data on the GP described in our methodology. In this figure we can see how results obtained with a target acceptance probability of $90\%$ are around $1.5$ times wider than those obtained with a target acceptance probability of $97 \%$ for which the number of divergences was reduced twenty-fold to around $0.5 \%$. divergent steps in our chains.

Finally, it is also important to mention that raising the target acceptance probability reduced the computational time required by the chains ran for this paper by an order of magnitude, proving once again the importance of tuning the numerical methods employed by the sampler.

% Don't change these lines
\bsp	% typesetting comment
\label{lastpage}
\end{document}